\documentclass[12pt,preprint]{aastex}
\usepackage{amssymb,amsmath,graphicx}
\setcitestyle{notesep={ }}

\begin{document}

\title{Disk wind feedback from high-mass protostars}
\author{Jan E. Staff}
\affil{College of Science and Math, University of the Virgin Islands, St
Thomas, 00802, United States Virgin Islands}
\author{Kei E. I. Tanaka}
\affil{Department of Earth and Space Science,Osaka University, Toyonaka, Osaka 560-0043, Japan\\and\\
ALMA Project, National Astronomical Observatory of Japan, Mitaka, Tokyo 181-8588, Japan}
\author{Jonathan C. Tan}
\affil{Department of Space, Earth \& Environment, Chalmers University of Technology, Gothenburg, Sweden\\ and\\ Department of Astronomy, University of Virginia, Charlottesville, Virginia, USA}

\date{}

\begin{abstract}
We perform a sequence of 3D magnetohydrodynamic (MHD) simulations of
the outflow-core interaction for a massive protostar forming via
collapse of an initial cloud core of $60~{M_\odot}$.  This allows us
to characterize the properties of disk wind driven outflows from
massive protostars, which can allow testing of different massive star
formation theories. It also enables us to assess quantitatively the
impact of outflow feedback on protostellar core morphology and overall
star formation efficiency.  We find that the opening angle of the flow
increases with increasing protostellar mass, in agreement with a
simple semi-analytic model.  Once the protostar reaches
$\sim24~{M_\odot}$ the outflow's opening angle is so wide that it has
blown away most of the envelope, thereby nearly ending its own
accretion.  We thus find an overall star formation efficiency of
$\sim50\%$, similar to that expected from low-mass protostellar
cores. Our simulation results therefore indicate that the MHD disk
wind outflow is the dominant feedback mechanism for helping to shape
the stellar initial mass function from a given prestellar core mass
function.
\end{abstract}

\section{Introduction}

Bipolar jets and outflows are commonly observed from accretion disks
around low-mass protostars
\citep[e.g.,][]{2000ApJ...537L..49B,2007prpl.conf..231R,2008ApJ...689.1112C}.
The launching of this outflow is thought to be due to
magnetocentrifugal acceleration
\citep{1982MNRAS.199..883B,2000prpl.conf..759K}, in which a
large-scale magnetic field threads the accretion disk.  Gas can flow
along the magnetic field lines if they are inclined sufficiently with
respect to the disk.  The gas gains speed as it flows along the field
lines.  Beyond the Alfv{\'e}n surface, the field lines will become
twisted, which collimates the flow.  Although typically more difficult
to observe, high-mass protostars are also often seen to have
associated jets and outflows
\citep[e.g.,][]{2007prpl.conf..245A,2016A&ARv..24....6B,2017NatAs...1E.146H}.  
Indeed,
outflows are very commonly seen in most astrophysical settings where
there is an accretion disk surrounding a central object, and the
magnetocentrifugal model was first proposed for AGN jets. The disk
wind mechanism has been studied extensively with numerical simulations
\citep[e.g.,][]{1985PASJ...37...31S,1985PASJ...37..515U,1997ApJ...482..712O,1997ApJ...482..708R,1997Natur.385..409O,2003ApJ...582..292O,2006ApJ...653L..33A,2009A&A...507.1203M,2010ApJ...722.1325S,2015MNRAS.446.3975S,2011ApJ...728L..11R,2014MNRAS.439.3641S}.
Other models for launching the outflow have also been proposed. For
instance, an outflow may originate in the innermost part of the disk
or the disk/magnetosphere boundary \citep[often referred to as the
  X-wind model,][]{1991ApJ...379..696L,2000prpl.conf..789S}, a stellar
wind \citep{2005ApJ...632L.135M}, or driven by the magnetic pressure
of the magnetic field \citep[i.e., magnetic tower model
  of][]{1996MNRAS.279..389L}.

One possible formation scenario for high-mass stars is that of Core
Accretion, i.e., it is simply a scaled-up version of the standard
model for low-mass star formation by accretion from gravitationally
bound cores (Shu, Adams \& Lizano 1987).  In the Turbulent Core Model
\citep{2002Natur.416...59M,2003ApJ...585..850M}, a combination of
turbulence and magnetic pressure provide most of the support in a
massive core against gravity. In high pressure conditions typical of
observed massive star forming regions, the accretion rate from such
massive cores is expected to be relatively high, i.e., with
$\sim10^{-4}\:$ to $\sim10^{-3}\:M_\odot\:{\rm yr}^{-1}$, compared to
lower-mass protostars in lower pressure regions, i.e., with
$\sim10^{-6}\:$ to $\sim10^{-5}\:M_\odot\:{\rm yr}^{-1}$. In this
scenario, the outflows from forming massive stars may therefore also
be a scaled-up version of the outflows from lower-mass stars, but with
higher mass outflow rates and momentum rates.
Alternative formation scenarios include models in which multiple
smaller objects form close together, and then collide to form larger
stars \citep{1998MNRAS.298...93B}, and Competitive Accretion
\citep{2001MNRAS.323..785B}, in which stars forming in central, dense
regions of protoclusters accrete most of their mass from a globally
collapsing reservoir of ambient clump material 
\citep[see][for a review]{2014prpl.conf..149T}.
In these models, outflows are expected to be more disordered.

There are some observations of highly collimated jets from massive
young stellar objects (YSOs). For example, \citet{1993ApJ...416..208M}
found a bipolar jet from the central source between HH 80 and 81. 
\citet{2018arXiv180108147M} reported observations of
HH 1177, a jet originating from a massive YSO in the Large Magellanic
Cloud.
\citet{2015A&A...573A..82C} observed jets from 18 intermediate
and high mass YSOs, and found that these jets appear as a scaled-up
version of jets from lower-mass YSOs.  \citet{2010Sci...330.1209C} and
\citet{2015A&A...583L...3S} studied the magnetic field morphology near
massive YSOs, and found a magnetic field configuration parallel to the
outflow and perpendicular to the disk.  Observations of wider angle,
but still collimated, molecular outflows have also been reported: see
for instance
\citet{2002A&A...383..892B,2004A&A...426..503W,2013ApJ...767...58Z,2014ApJ...792..116Z,2015MNRAS.453..645M}.
The general trend found in these works is that a more luminous (and
hence generally more massive) protostar has more massive and powerful
outflows, perhaps with larger opening angles.

The collapsing gas in a core may be dispersed by the outflows and jets
coming from forming stars \citep[e.g.,][]{2000ApJ...545..364M}.  This
occurs both because some gas is ejected from the accretion disk into
the outflow, and also because the outflow sweeps up gas in the core as
it propagates outwards.  
If the opening angle of the outflow is small, not much gas is being
swept up, while an outflow with a large opening angle will sweep up more gas.
This feedback on the core can therefore
regulate the core to star formation efficiency (SFE) and this can be related to the opening
angle of the flow, which is an observable quantity.  
We denote the SFE by $\epsilon_{\rm core}$ and defined it to be the final mass
of the star divided by the initial core mass.  Understanding the SFE
can allow for the transformation of the prestellar core mass
function (CMF) to the stellar initial mass function (IMF).

\citet{2014ApJ...788..166Z} performed semi-analytic modeling and
radiative transfer calculations of massive 
protostars forming from
massive cores, based on the turbulent core model and including MHD disk
wind outflow feedback.  They found that a $60\:M_\odot$ core resulted
in a $26\:M_\odot$ star, i.e., a SFE of $\sim43\%$.
\citet{2016ApJ...832...40K} performed axisymmetric radiation 
hydrodynamic (HD)
simulations of $100\:M_\odot$ cores with a subgrid module for protostellar outflow feedback, and found SFEs of
$\sim20\%-50\%$.  
Using semi-analytic models extended from those of
\citet{2014ApJ...788..166Z}, \citet{2017ApJ...835...32T} studied
feedback during massive star formation, and found the disk wind to be
the dominant feedback mechanism, with overall SFEs of $\sim30-50\%$.
\citet{2012MNRAS.421..588M} investigated the SFE in low-mass cores by
doing resistive magnetohydrodynamic (MHD) simulations, and found a SFE of $\lesssim50\%$ in
those cases.  \citet{2017MNRAS.470.1026M,2018MNRAS.475..391M} presented results of MHD simulations of
outflows from massive YSOs.  Their results indicated that massive
stars can form through the same mechanism as low mass stars, though
they did not follow the evolution until the end, and therefore could
not estimate the SFE. Recently, \citet{2018A&A...620A.182K} performed axisymmetric, 
non-ideal MHD collapse simulations of a $100~{\rm M_\odot}$ core, and 
followed the evolution until the protostar reached a mass of 
$\sim70~{M_\odot}$.

Observationally, \citet{2010A&A...518L.106K} and
\citet{2010A&A...518L.102A} reported that in relatively low mass
clusters, the CMF and IMF have similar shapes, but the CMF is shifted
to higher masses by a factor a few. \citet{2018ApJ...853..160C} have measured
the CMF in a more massive protocluster finding a similar shape as the
Salpeter IMF, which may indicate that SFE is relatively constant with
core mass. However, \citet{2018ApJ...862..105L} and \citet{2018NatAs...2..478M}
have
claimed shallower, i.e., top-heavy, high-end CMFs, which may imply
SFEs become smaller at higher masses, potentially consistent with the
results of \citet{2017ApJ...835...32T}.

Here we present results of three dimensional ideal MHD simulations of
the outflow from a protostar, forming from an initial core of
$60\:M_\odot$. With these MHD simulations, we aim to test the semi-analytic modeling of \citet{2014ApJ...788..166Z}. We describe the method in \S2. We present our results
in \S3, and discuss them in \S4 where we also summarize our results.

\section{Methods}
\label{methodssection}

\subsection{Overview}

We consider the formation of a single massive star from the collapse
of a cloud core under the framework of the Turbulent Core Model
\citep{2003ApJ...585..850M}. The initial mass of the core, which is a 
basic parameter of the model, is here taken to be
$M_c=60\:M_\odot$. The core is assumed to be in pressure equilibrium 
with an ambient self-gravitating clump environment, which is characterised by
its mass surface density---a second basic parameter of the model.
Typical observed values of mass surface densities of clumps that form high-mass 
stars are about $1\:{\rm g\:cm^{-2}}$, which we adopt for the case simulated here.
This sets the bounding pressure on the core and thus a core radius of $R_{c}=0.057$~pc 
or $\simeq 12,000\:{\rm au}$ \citep{2003ApJ...585..850M}. With the overall mean density
of the core set by these parameters, its collapse time to form a star is about 100,000 years.
The infalling material is assumed to join a central disk, through which gas accretes onto the central protostar.  The accretion process drives a disk wind via the magnetocentrifugal mechanism \citep{1982MNRAS.199..883B}, creating a powerful outflow that reduces the
infall rate and the SFE from the core.  
A poloidal magnetic field threads the core
with a total initial magnetic flux in the core of $1~{\rm mG}\times R_c^2$, similar to the value of the fiducial model of \citet{2003ApJ...585..850M}.
To investigate the properties
of the outflow and the SFE from the core, we perform 3-D MHD simulations
using the ZEUS-MP code \citep{2000RMxAC...9...66N}.  

However, to follow the full process of star formation from start to finish over the long time period of the collapse of the core, i.e., $\sim 10^5$~yr,
while at the same time resolving the disk, especially the inner disk, and its launching of an outflow, is computationally extremely expensive because of
the very different scales involved. In order
to carry out a practical computation, we make two simplifications:
1) Instead of simulating the entire long-term evolution, we divide the
problem into an evolutionary sequence of models with protostellar masses 
$m_*=1,\:2,\:4,\:8,\:16$, and 
$24\:{M_\odot}$ and simulate these for relatively short periods, assuming they are quasi steady
states. 2) To avoid the extremely high resolution needed to properly
resolve the launching of the wind from the disk, we instead inject
a disk wind from the boundary of the simulation box set to be 100 au above the midplane (see 
Figure~\ref{gridlayoutfig} for a schematic illustration).
With these simulations, we will then examine the quasi
equilibrium behavior of the system, especially the opening angle
($\theta_{\rm outflow}$) of the outflow cavity, which helps determines the SFE
from the core.
The accretion rate giving the power of the wind, and the disk radius
is taken from the semi-analytic model described by \citet{2014ApJ...788..166Z}. 
Density and velocity profiles for the injected disk wind are taken from 
\citet{2015MNRAS.446.3975S}, who performed high resolution simulations
of the jet/wind from the disk surface out to $\sim 100~{\rm au}$, which helps determine
our choice of the height of the the injection boundary
in our simulations to be this value
(see Figure~\ref{gridlayoutfig}).

At each stage of the sequence, the
protostellar disk is assumed to be massive, i.e., the disk mass is a constant fraction $f_d=1/3$ of the protostellar mass, and thus possibly moderately
self-gravitating due to the high mass supply from the infalling
envelope.  
The disk and stellar radii are held constant for each model, but change from model to model in the evolutionary sequence,
following \citet{2014ApJ...788..166Z} (see Table~\ref{setupparamtable}).
As a first simple approach, we initiate each model in the sequence with 
a spherically symmetric core, without an outflow cavity produced by
the earlier outflow.
To test this approximation, once the opening angle
is seen to become significant, we also run a sequence of models with
$m_*=4,\:8,\:16$, and $24\:{M_\odot}$ where the initial setup has a
``pre-cleared'' cavity mimicking the effect of the outflow earlier in
the evolution. 
We describe this pre-cleared cavity in more detail in section~\ref{preclearedsection}.

We run each simulation for an amount of time needed for the star to accrete half of the mass needed to bring it to the next simulated model based on the analytic accretion rates of the Turbulent Core Model. For example, we run the $8~{M_\odot}$ simulation until it would have accreted $4~{M_\odot}$, which is roughly $15,000~{\rm years}$. However, for the $24~{M_\odot}$ case, which is near the end of the formation process, we run the simulation for $\sim 12,000$ years, i.e., until it would have accreted $\sim4~{M_\odot}$.
The accretion rates vary between 
$1.0-3.3\times10^{-4}~{\rm M_\odot\:{\rm yr}^{-1}}$, following the estimates
in \citet{2014ApJ...788..166Z}, see Table~\ref{setupparamtable}.

\subsection{Grid setup}

We use Cartesian coordinates
($x_1$, $x_2$, $x_3$) to describe our domain, which contains most of
one hemisphere, i.e., $100~{\rm au}<x_1<R_c+\zeta_1$,
$-R_c-\zeta_2<x_2<R_c+\zeta_2$, and $-R_c-\zeta_3<x_3<R_c+\zeta_3$, with
$\zeta_1$ and $\zeta_2=\zeta_3$ ensuring that the boundary is outside of the
core. This is illustrated in Figure~\ref{gridlayoutfig}.  
The exact values of $\zeta_1$, $\zeta_2$, and $\zeta_3$ depend on the
simulation.  In order to be able to cover the entire core-scale on the
grid, while maintaining a reasonable resolution near the injection region, 
we use
a Cartesian coordinate system with logarithmically spaced grid cells
(``ratioed'' grid in ZEUS terminology).  This means that in the $x_1$
direction, the grid cells are fairly small  
near the inner $x_1$
boundary, and gradually become larger farther away from this boundary (see below for details).
In the $x_2$ and $x_3$ directions, the grid cells are fairly small
near the central axis, and gradually become larger farther away from
the axis.  As a consequence, the grid cells can become rather large
and deviate substantially from a cubic shape in the outer regions of
the core.  To ensure that these rather coarse grid cells do not affect
the dynamics, we have also performed two comparison simulations with
higher resolution (see \S\ref{ressubsection}).

The number of grid cells varies between the simulations, with the $1$
and $2\:M_\odot$ simulations having more cells because the injection
region is relatively smaller compared to the core size than in the
higher mass simulations.  We aim at resolving the scale of the disk
wind injection region, $r_{\rm inj}$, with $\sim10$ cells across.
The disk radius increases with time (see Table~\ref{setupparamtable}). 
Thus we increase the 
injection radius for the higher protostellar masses in the sequence, leading to a change in the number of grid cells between the simulations.
The injection scale dictates how small the smallest cells
around the axis are.  There is a limit to how large a ratio between
one cell and the next ZEUS-MP will allow, which therefore sets a lower
limit on how many grid cells are needed in order to cover the whole
core-hemisphere.
These considerations lead us to use a grid with 
$210\times380\times380$ cells for the $m_*=1$ and $2~{M_\odot}$
simulations, and $140\times260\times260$ cells in our standard
setup for the simulations with $m_*\geq4~{M_\odot}$.
To test the effect of grid resolution on the results, we
also ran the $4~{M_\odot}$ simulation using a grid with
$210\times380\times380$ cells (medium resolution), and using a grid
with $280\times520\times520$ cells (i.e., double the number of cells;
high resolution).  

The inner $x_1$ boundary is a special boundary, where we assume that 
the density and
velocity are held constant at all times.
Here the boundary condition in ZEUS-MP is ``inflow''. 
To control the magnetic field on such a boundary, one can set the 
electromotive force (emf) there.
However, as discussed in the appendix of \citet{2003ApJ...582..292O}, it is 
unclear how to set the optimal boundary conditions in this case.
We therefore set the emfs to zero on this boundary.
All other boundaries are normal ZEUS outflow boundaries.

\subsection{Physical initial conditions for the evolutionary sequence}

The density structure of the prestellar core in the
fiducial Turbulent Core Model is assumed to be spherical,
following a power law of
the form $\rho \propto r^{-1.5}$ \citep[see][]{2003ApJ...585..850M}.
As the collapse starts, the density profile is expected to become
shallower.  Thus, based on the self-similar solution of inside-out
collapse \citep{1977ApJ...214..488S,1997ApJ...476..750M}, we 
approximate the initial condition for the density profile of the
envelope with a power law of index -1:
\begin{equation}
\rho_{\rm env}(t=0) = \rho_{\rm env,out} \left( r/R_c \right)^{-1},
\end{equation}
where $r$ is the distance from the stellar center, and $\rho_{\rm
  env,out}$ is a normalization density to give the appropriate total
mass of the envelope, i.e., $M_{\rm env}=M_c-(1+f_d) m_*$.  
The core radius is kept constant in all the models in the sequence,
i.e., $R_c=0.057\:$pc.
Although
the collapse has started, we assume initial velocities in the envelope
to be zero for simplicity.  
Assuming that the number density of helium nuclei $n_{\rm He}$ is $10\%$ of 
that of hydrogen nuclei $n_{\rm H}$ and ignoring the contributions of other elements, 
we set a mass per H nucleus of $2.34\times10^{-24}~{\rm g}$,
which corresponds to a mean molecular weight of $2.33$.
We approximate the gas as being
isothermal, with a temperature of $100~{\rm K}$ chosen to be representative of a massive protostellar core, giving a sound speed
$c_{s}=0.6~{\rm km~s^{-1}}$ for
molecular gas.

We include the gravitational potential from the protostar, which is
taken to be a point mass of $m_*$. For the infall envelope, for
simplicity we also treat this via a static gravitational potential
based on the initial gas mass distribution in the core, i.e., mass
$M_{\rm env}$.  Note that in this approximation the minor contribution
to the potential of the disk is ignored.  Tests show that this only
has minor effects on the results.

The core is threaded by a magnetic field, which has two contributions.
We expect the magnetic field of the core to be dragged along with the accreting
material towards the protostar, giving it an ``hour-glass shape''.
There is therefore a poloidal (``hour-glass shaped'') ``Blandford-Payne'' 
(BP) like force-free disk-field (Figure~\ref{gridlayoutfig} also shows 
a schematic illustrating this field) originating on the
accretion disk \citep[the poloidal magnetic field on the 
midplane scales as $r^{-1.25}$,][]{1982MNRAS.199..883B,
  2001A&A...379.1170J}.  
This disk field is normalized as in
\citet{2015MNRAS.446.3975S} (assuming equipartition at the inner edge
of the disk; which we assume extends all the way to the stellar surface), 
scaled to the relevant protostellar mass in each
simulation.  In addition, we add a uniform
field in the $x_1$ direction to this field everywhere, so that the total flux
of the initial core is $1~{\rm mG}\times R_c^2$.  The uniform
field dominates over the disk field in the outer regions of the core.
Near the star (and in the central region of our simulation box), the
uniform field is much weaker than the BP field.

We also run the $m_*=16\:M_\odot$ simulation with no magnetic field as
a test case.  Here, we keep the setup from the regular runs, and
simply set the magnetic field strength everywhere to zero.  The
outflow material is therefore injected in the same direction as in the
simulation with magnetic field (see below). This simulation helps to illustrate
the role played by the magnetic field.

\subsection{Injection of the disk wind}
\label{injmethodsection}

\begin{figure}
\includegraphics[width=0.9\textwidth]{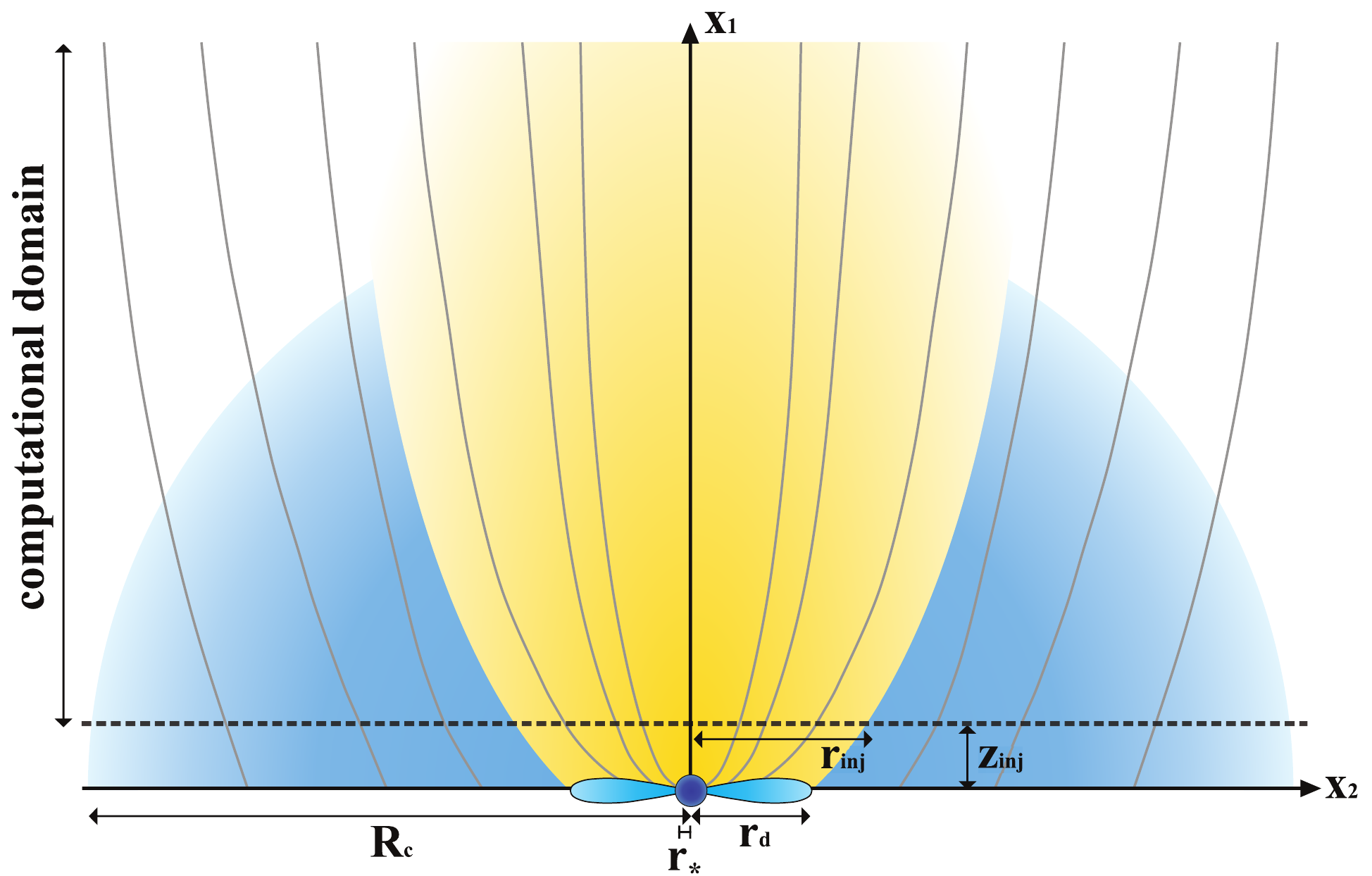}
\caption{
Schematic illustration of the simulation setup. The $x_3$ direction is
perpendicular to the displayed image, and shows similar features as in the $x_2$ direction. The zero-point on the $x_1$
axis is on the protostar.
Also shown is the disk,
the injection region, outflow, and core infall envelope. The gray
lines illustrate the shape of the initial magnetic field lines.}
\label{gridlayoutfig}
\end{figure}

One of our objectives is to test the semi-analytic modeling of
\citet{2014ApJ...788..166Z}, and we therefore ensure that the mass flow
and momentum rates are similar to those in that work.
Our injection boundary is at a height of
$100~{\rm au}$ above the disk (see Figure~\ref{gridlayoutfig}).
For the density and velocity profiles, we adopt the results of the 
``BP'' MHD simulations in \citet{2015MNRAS.446.3975S}.
They simulated the driving process of the
jet/outflow from the disk surface on scales of $\lesssim100~{\rm au}$ for
a low mass protostar. Those were ideal MHD simulations, thus fully 
scalable for the protostellar mass and radius 
\citep[see also][]{1997ApJ...482..712O}.
In this subsection, we outline the boundary conditions of this injected disk wind.
The setup parameters for the simulations
are summarized in Table~\ref{setupparamtable}.

The width of the flow at the injection radius
(a height of $100~{\rm au}$) has been calculated based on the shape
of the field lines in the ``Blandford-Payne'' magnetic field
configuration \citep[eq. B22 in][]{2013ApJ...766...86Z}:
\begin{equation}
\frac{r_{\rm inj}}{r_{d}}=1+14\ln\bigg(1+0.07\frac{z_{\rm inj}}{r_{d}}\bigg),
\end{equation}
where $r_{d}$ is the disk radius from \citet{2014ApJ...788..166Z}.

There are three contributions to the injection velocity.
The injection velocity profile along the $x_1$ direction
is found by fitting a power law to the results in \citet{2015MNRAS.446.3975S}:
\begin{equation}
v_{\rm inj}=(r_{\rm cyl}/r_*)^{-1/2}\phi_{\rm inj} v_{\rm K*},
\label{vinjeq}
\end{equation}
where $r_*$ is the radius of the protostar, $v_{\rm K*}$ is the 
Keplerian velocity at the stellar surface,
$\phi_{\rm inj}$ is a dimensionless factor to ensure that
the injected mass and momentum rates are equal to those of
\citet{2014ApJ...788..166Z}, and $r_{\rm cyl}=\sqrt{x_2^2+x_3^2}$ is the distance from the axis.
Hence we used the velocity profile obtained from the simulations in 
\citep{2015MNRAS.446.3975S}, and scaled it to ensure the mass and the momentum
rates are as found in \citet{2014ApJ...788..166Z}. We find that $\phi_{\rm inj}$ takes values between 40 and 100, i.e., relatively large values due to the resolution constraints of our numerical simulation grid.

The injected velocity is given additional
components in the $x_2$ and $x_3$ directions to angle it in the same
direction as the initial magnetic field.
The magnitudes of these
depend on the inclination of the field lines, which have an angle between
$\sim50^\circ$ and $\sim90^\circ$ (with respect to the disk-plane) in the
injection region, i.e., this
additional poloidal component is less than, but can be comparable to, the vertical injection
speed.

In order for the injected flow to be rotating it is given an additional toroidal
velocity component:
\begin{equation}
v_{\rm \phi, inj}=0.23 \bigg(\frac{r_{\rm cyl}}{22.4 r_*}\bigg)^{-1/2}v_{\rm K*}.
\label{vphiinjeq}
\end{equation}
This expression was found by fitting a power law to the toroidal velocity found by \citet{2015MNRAS.446.3975S}.
The injected speed and direction is kept constant throughout the
simulation.  The toroidal (rotational) speed is only a few percent of
the poloidal speed, so this only leads to a small deviation from the
initial field direction.  As the magnetic field evolves throughout the
simulation, the deviation may however change at later times.

The density of the injected disk wind material has also been found based on the results of \citep{2015MNRAS.446.3975S}, and is given by
\begin{equation}
\rho_{\rm inj}=\begin{cases}
\exp{(0.0289~r_{\rm cyl}/r_*)}\phi_{\rho} \rho_{\rm 0} & \quad r_{\rm cyl}<x_0\\
2.77\bigg(\dfrac{r_{\rm cyl}}{x_0}\bigg)^{-1}\phi_{\rho} \rho_{\rm 0} & \quad r_{\rm cyl}\ge x_0\\
\end{cases}
\label{rhoinjeq}
\end{equation}
with $x_0=35.3  r_*$ and $\rho_{\rm 0}$ being the injection density on
the axis, which is set to match the accretion rate from
\citet{2014ApJ...788..166Z} and by assuming that the injected mass flux
is $10\%$ of the accreted mass flux.  Such a fiducial ratio of mass
outflow rate to accretion rate is consistent with observational
estimates \citep[e.g.,][]{2002A&A...383..892B,2016A&ARv..24....6B},
although these are quite uncertain.  We note that in our current simulations
the resolution is larger than $x_0$, so we only use the second line in
equation~\ref{rhoinjeq}.
In the $4~M_\odot$ simulation with the highest resolution, the finest
cell size is $50 r_*$.

The density profile of the injected material in 
\citet{2014ApJ...788..166Z} is assumed to be
$\rho_{\rm inj}\propto r_{\rm cyl}^{-1.5}$ at the disk.
At a height of $100~{\rm au}$ in that work, the density profile
of the outflow
has a power of about $-1.25$ for the innermost $15~{\rm au}$, but
deviates substantially from a power law at larger radii.
This is to be compared with the $-1$ power law in our simulations, based on the work of \citet{2015MNRAS.446.3975S}.
It is because of this
difference, and because we have limited resolution in our simulations,
that we need the factor $\phi_{\rho}$ (typically around 0.5-1 depending on the simulation) in the expression for $\rho_{\rm
  inj}$, and $\phi_{\rm inj}$ in the expression for $v_{\rm inj}$ in order
to obtain the same mass flow and momentum rate as in
\citet{2014ApJ...788..166Z}.

\begin{rotate}
\begin{table}
\caption{
Summary of the setup parameters for the simulations. $m_{*}$ is
	the mass of the protostar, $\dot{m}_*$ is the accretion rate
	onto the protostar, $r_{*}$ is the radius of the protostar,
$r_{d}$ is the radius of the disk,
$r_{\rm inj}$ is the radius of the injection region,
$v_{\rm K*}$ is the Keplerian speed at the protostellar surface,
$\rho_{\rm 0}$ is the injection density on the axis, $\rho_{\rm env, out}$ is
the density at the core boundary,
$B_{*}$ is the disk magnetic field at the surface of the
	protostar, $\dot{m}_{\rm inj}$ is the rate of outflow mass 
	injection (into one hemisphere; equal to 5\% of $\dot{m}_*$), and
$\dot{p}_{\rm inj}$ is the rate of momentum injection (into one hemisphere).}
\begin{tabular}{ccccccccccccc}
\hline
	$m_{*}$ & $\dot{m}_*$ & $r_{*}$ & $r_{d}$ & $r_{\rm inj}$ & $v_{\rm K*}$ & $\rho_{0}$ & $\rho_{\rm env,out}$ & $B_{\rm *}$ & $\dot{m}_{\rm inj}$ & $\dot{p}_{\rm inj}$ & \\
	$[M_{\odot}]$ & $[M_\odot~{\rm\:yr^{-1}}]$ & $[R_{\odot}]$ & $[{\rm au}]$ & $[{\rm au}]$ & $[{\rm km~s^{-1}}]$ & $[{\rm cm^{-3}}]$ & $[{\rm cm^{-3}}]$ & $[{\rm G}]$ & $[M_\odot~{\rm\:yr^{-1}}]$ & $[M_\odot~{\rm\:yr^{-1}}~{\rm km~s}^{-1}]$\\
	$1$ & $1.01\times10^{-4}$ & 2.61 & 13.1 & 91.9 & 270.31 &$2.9\times10^{8}$ & $1.6\times10^{6}$ & 324.35 & $5.05\times10^{-6}$ & $0.0025$ \\
	$2$ & $1.42\times10^{-4}$ & 3.45 & 20.9 & 105.7 & 332.50 & $1.7\times10^{8}$ & $1.6\times10^{6}$ & 305.58 & $7.10\times10^{-6}$ & $0.0050$ \\
	$4$ & $1.95\times10^{-4}$ & 20.5 & 34.2 & 123.7 & 192.90 & $2.3\times10^{7}$ & $1.5\times10^{6}$ & 68.152 & $9.75\times10^{-6}$ & $0.0047$ \\
	$8$ & $2.67\times10^{-4}$ & 33.4 & 57.1 & 150.0 & 213.73 & $1.1\times10^{7}$ & $1.3\times10^{6}$ & 51.105 & $1.34\times10^{-5}$ & $0.0071$ \\
	$16$ & $3.19\times10^{-4}$ & 6.41 & 101.0 & 196.4 &689.95 & $3.3\times10^{7}$ & $1.0\times10^{6}$ & 271.99 & $1.60\times10^{-5}$ & $0.021$ \\
	$24$ & $3.34\times10^{-4}$ & 6.38 & 185.0 & 282.0 & 848.6 & $2.0\times10^{7}$ & $7.4\times10^{5}$ & 262.83 & $1.67\times10^{-5}$ & $0.025$ \\
\hline
\end{tabular}
\label{setupparamtable}
\end{table}
\end{rotate}

\subsection{Calculating the outflow opening angle $\theta_{\rm outflow}$}

We are particularly interested in the opening angle ($\theta_{\rm
  outflow}$) of the outflow, since this is a quantity that is directly
related to the star formation efficiency and is a measurable quantity
in real protostellar cores.  It was also calculated by
\citet{2014ApJ...788..166Z}, and can therefore be compared with their
work.  
To calculate it, we search through the
grid (at $x_1=R_{\rm c}$) for forward velocities 
$v_1>c_s=0.6~{\rm km~s^{-1}}$. Starting from 
the outside grid boundary, we seek through the grid along the 
principal axes towards the central rotation axis, and take the
first instance of such velocity to be the edge of the flow.  We then
draw a straight line from there to the protostar (the center of the
core).  The angle that this line makes with the normal to the disk we
define to be $\theta_{\rm outflow}$.  As the flow is not entirely symmetric,
we do this along both the $x_2$ and the $x_3$ directions, from both
boundaries, and take $\theta_{\rm outflow}$ to be the
average of these.
If the opening angle is so large that the outflow 
escapes through the side boundaries, then the angle is calculated at
the height at which it escapes, using a similar procedure as described above.

\subsection{Cases with a pre-cleared cavity}
\label{preclearedsection}

As described above, our first set of simulations ignore the earlier
evolutionary stages in the development of the outflow cavity.  To
explore the potential effects of this approximation, we run another
sequence of models with $m_*=4,\:8,\:16$, and $24\:{M_\odot}$, with
each model having a cavity pre-cleared based on the results of the
lower mass model without pre-clearing. 
In the pre-cleared region, we
simply reduced the density by a factor of ten compared to the
simulation without pre-clearing.  The boundary of the pre-cleared region was defined
to be:
\begin{equation}
r_{\rm cyl} = r\sin\theta_{\rm clearing} + 50 \left(\frac{R_\odot}{r_*}\right) {\rm au}
\end{equation}
where $\theta_{\rm clearing}$ is the opening angle $\theta_{\rm outflow}$ 
of the previous lower-mass simulation (Table~\ref{openingangletable}), 
and $r_{\rm cyl}$ and r are, as before, the cylindrical distance from 
the axis and the distance from the protostar.
This
expression was found to mimic the cavity found in the simulations
without pre-clearing reasonably well. 

\section{Results}

\subsection{General outflow morphologies and velocity distributions}

\begin{figure}
\includegraphics[width=\textwidth]{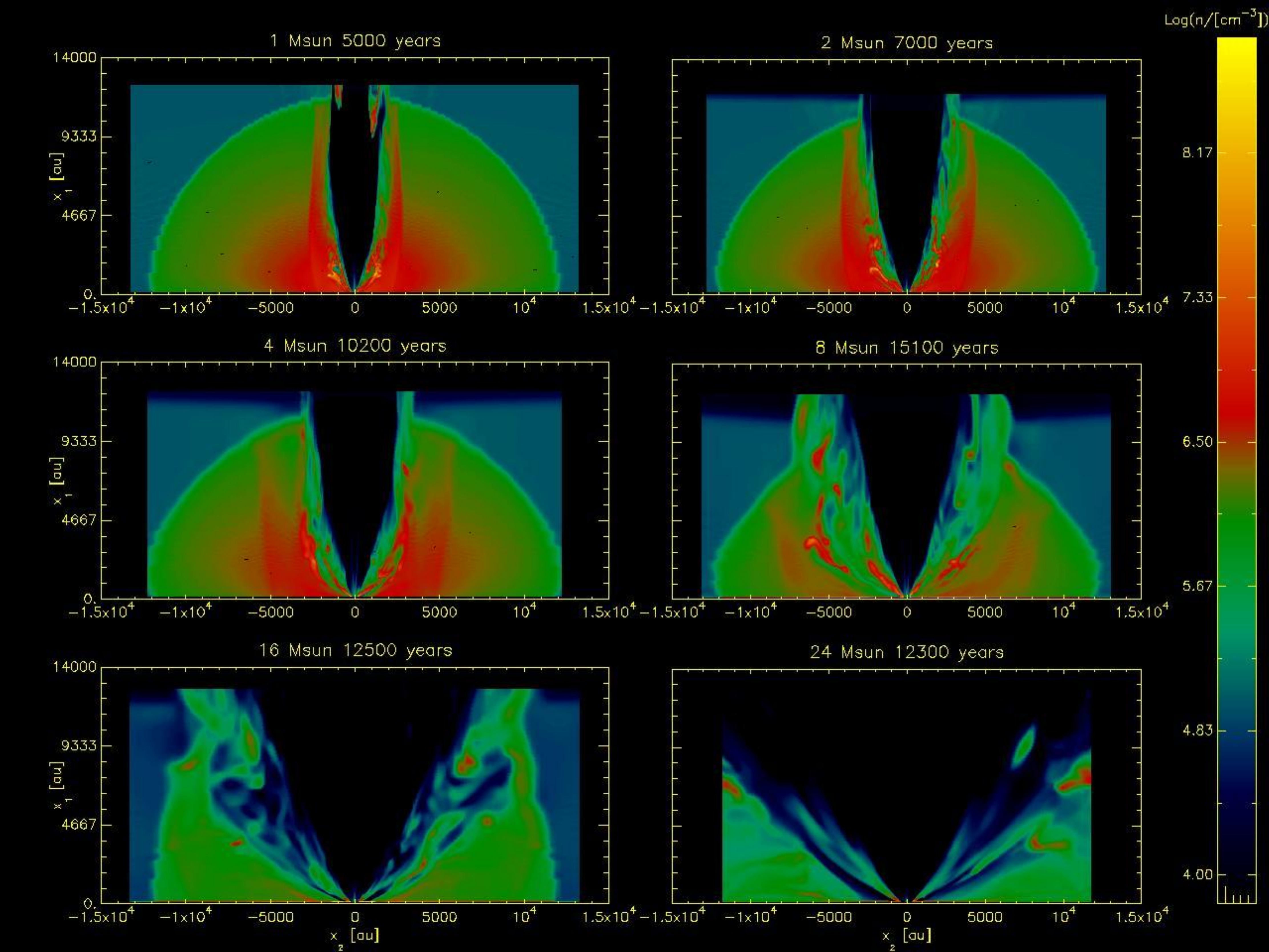}
\caption{
Slices through the middle of the simulation domains 
(in the $x_1$ - $x_2$ plane) showing density
structures ($n_{\rm H}$) of the massive protostellar cores for
fiducial runs with protostellar masses of 1, 2, 4, 8, 16, and 24
${M_\odot}$, as labelled. Outputs are shown after various amounts of
time evolution (see text).}
\label{allresultsr}
\end{figure}

\begin{figure}
\includegraphics[width=\textwidth]{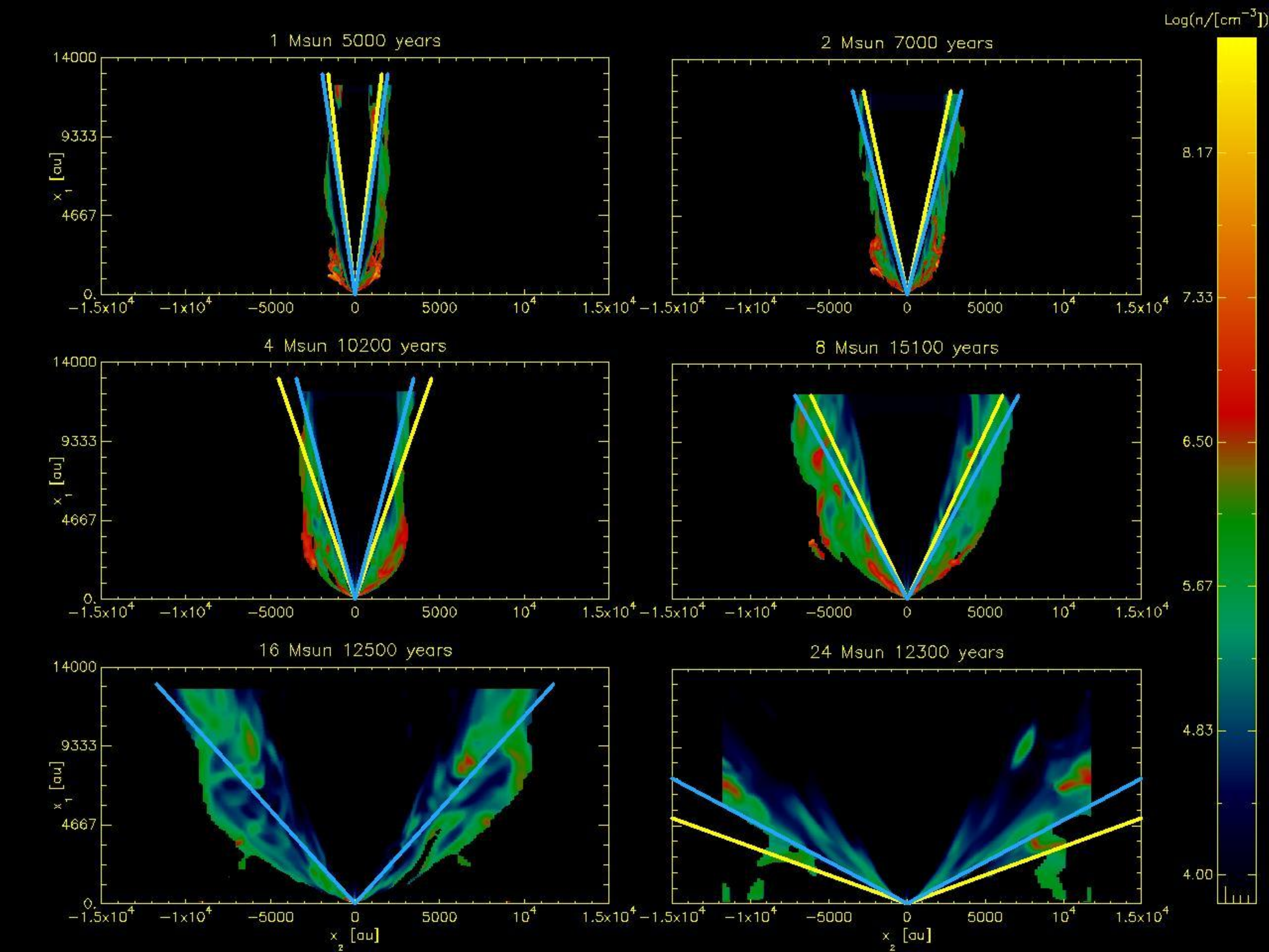}
\caption{
The same as Fig.~\ref{allresultsr}, but only showing the density
that has a positive velocity in the $x_1$ direction greater than the
sound speed ($v_1>c_s$).  This therefore
shows the structure of the outflow.  The yellow ``V''-shaped lines show
$\theta_{\rm outflow}$ found in the semi-analytic models of
\citet{2014ApJ...788..166Z}, while the light blue ``V''-shaped lines show 
the opening angle that we find in this work. Note that in the $16~{M_\odot}$ simulation, the blue and yellow lines are on top of each other, as the opening angle that we found matches that from \citet{2014ApJ...788..166Z}.}
\label{allresultsposr}
\end{figure}

We perform a sequence of disk wind protostellar outflow simulations
with the protostellar mass increasing from 1 to 24~$M_\odot$ and the
initial envelope mass declining from $59~{M_\odot}$ to 
$28~{M_\odot}$, which maintains the constant total of $60\:M_\odot$
(of the envelope, the protostar, and the disk; see above).  
We show the density slices at $x_3=0$ in Figure~\ref{allresultsr}
at the end
of the simulations.
Figure~\ref{allresultsposr} shows the same,
but only showing the material with a velocity component $v_1>c_s$.
Overlayed on that are yellow lines showing
the opening angle as found in \citet{2014ApJ...788..166Z}, and blue
lines showing the opening angle that we find in this work. 
As we discuss in more detail below, we find a good agreement between the
opening angle in this work and the analytic calculations of 
\citet{2014ApJ...788..166Z}.
In all
simulations, the outflow carves out a low density cavity.  However,
some higher density gas outside of this cavity is also outflowing, and
$\theta_{\rm outflow}$ is therefore larger than just the size of the
cavity.  As described in \S\ref{methodssection}, the time of the 
snapshots shown in Figure~\ref{allresultsr} is
after an amount of time needed for the star to accrete half of the
mass needed to bring it to the next simulated model.

\begin{figure}
\includegraphics[width=0.99\textwidth]{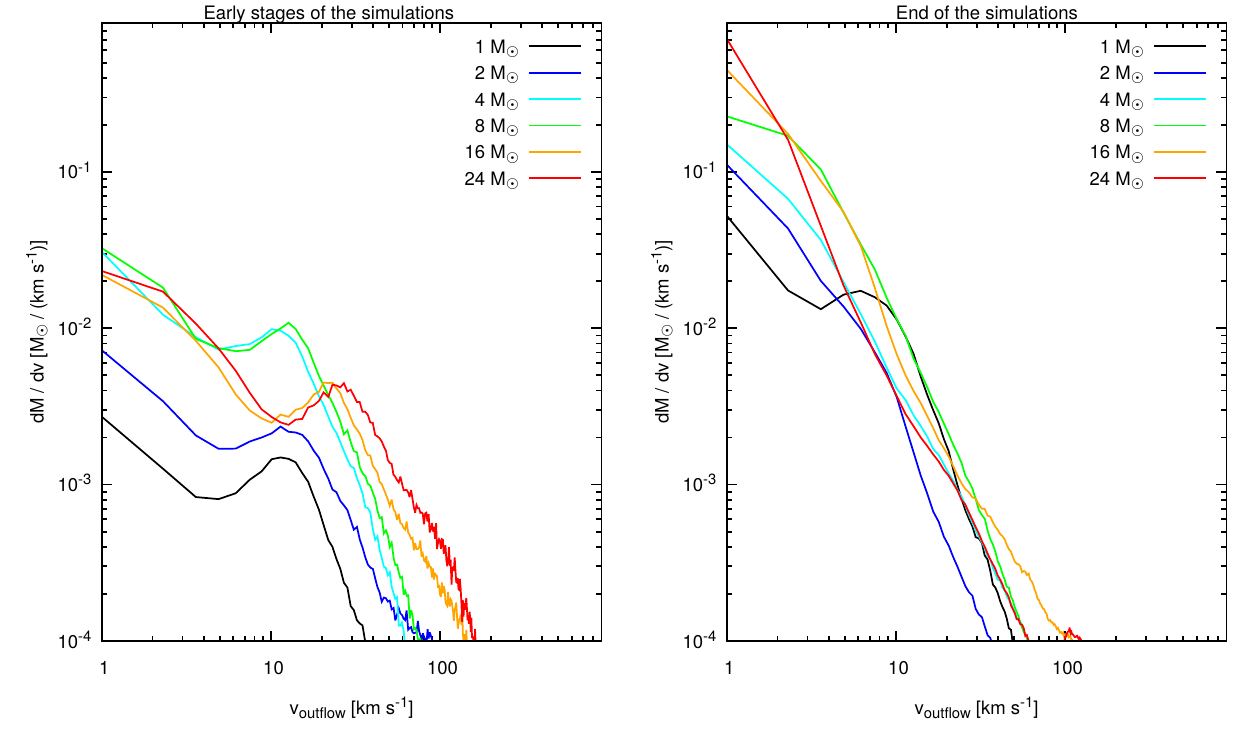}
\caption{
{\it Left panel:} Histogram showing the distribution 
of the outflow mass with respect to the outflow speed, evaluated around the time that the outflow breaks out of the core in each simulation. The bin width is
$1.3~{\rm km~s^{-1}}$. 
{\it Right panel: } the same, but at the end of the simulations, at which point much of the original mass has left the computational domain.}
\label{histograms}
\end{figure}

In Figure~\ref{histograms} we show histograms of the distribution of the
outflowing mass from one hemisphere with respect to the outflow velocity ($v_1$), first at an early stage of the 
simulations near the point of first break out from the core, and then at the end of
the simulations. In the former, the
lower mass simulations show a local peak around $10~{\rm km~s^{-1}}$, while
in the higher mass cases this peak rises to around $30~{\rm km~s^{-1}}$. We will compare these distributions with observed systems below.

\begin{figure}
\includegraphics[width=0.99\textwidth]{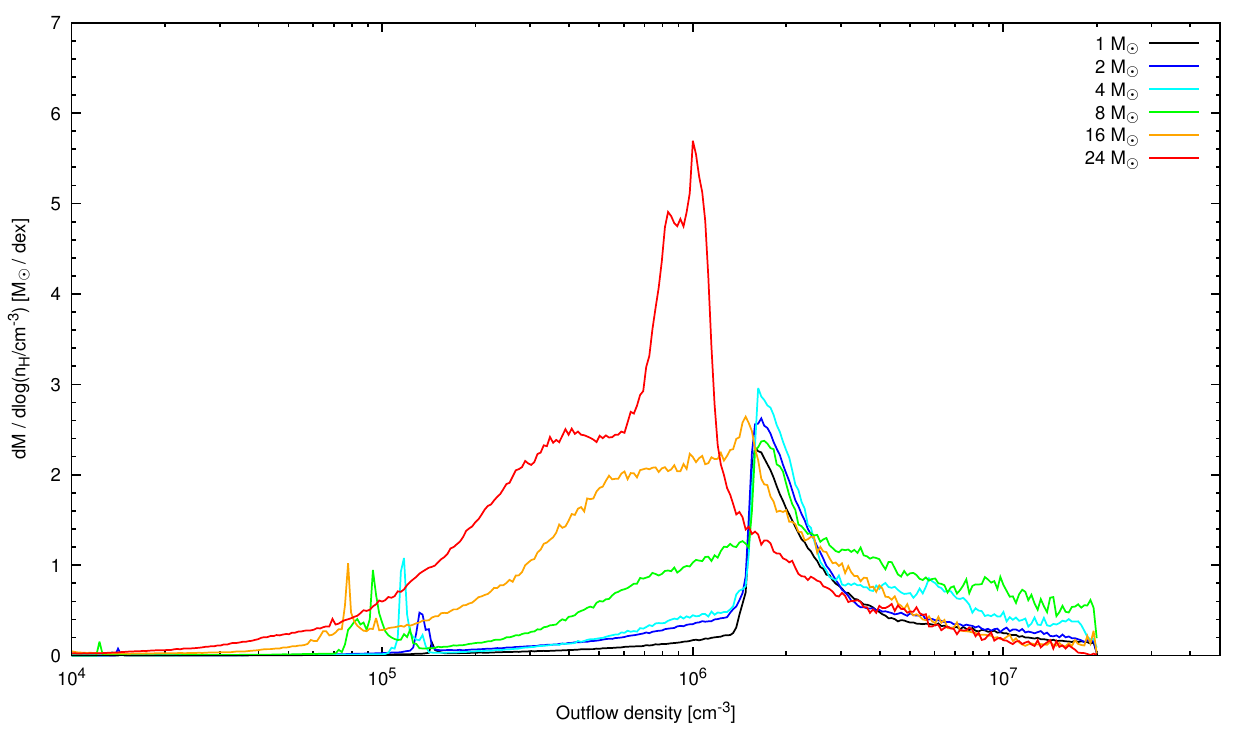}
\caption{
Distribution of the outflow mass with respect to the
logarithm of H number density. The histogram is made from the last
snapshot of each simulation. The bin width is 0.01 dex.}
\label{histogramsd}
\end{figure}

In Figure~\ref{histogramsd} we show the distribution of the outflowing
mass with respect to the outflow density, at the end of each
simulation. Most outflowing mass in the $24~{M_\odot}$ simulation
is found to be around $n_{\rm H}\sim 10^5-10^{6}~{\rm cm^{-3}}$. For the
lower-mass simulations, the distribution is bimodal with most mass at a 
density of around $2\times10^{6}~{\rm cm^{-3}}$ and 
another, smaller peak at $\sim10^5~{\rm cm^{-3}}$. The $16~{M_\odot}$
simulation is in between, with a much broader peak stretching from
$\sim5\times10^4$ to $\sim 10^6~{\rm cm^{-3}}$.

\begin{figure}
\includegraphics[width=0.99\textwidth]{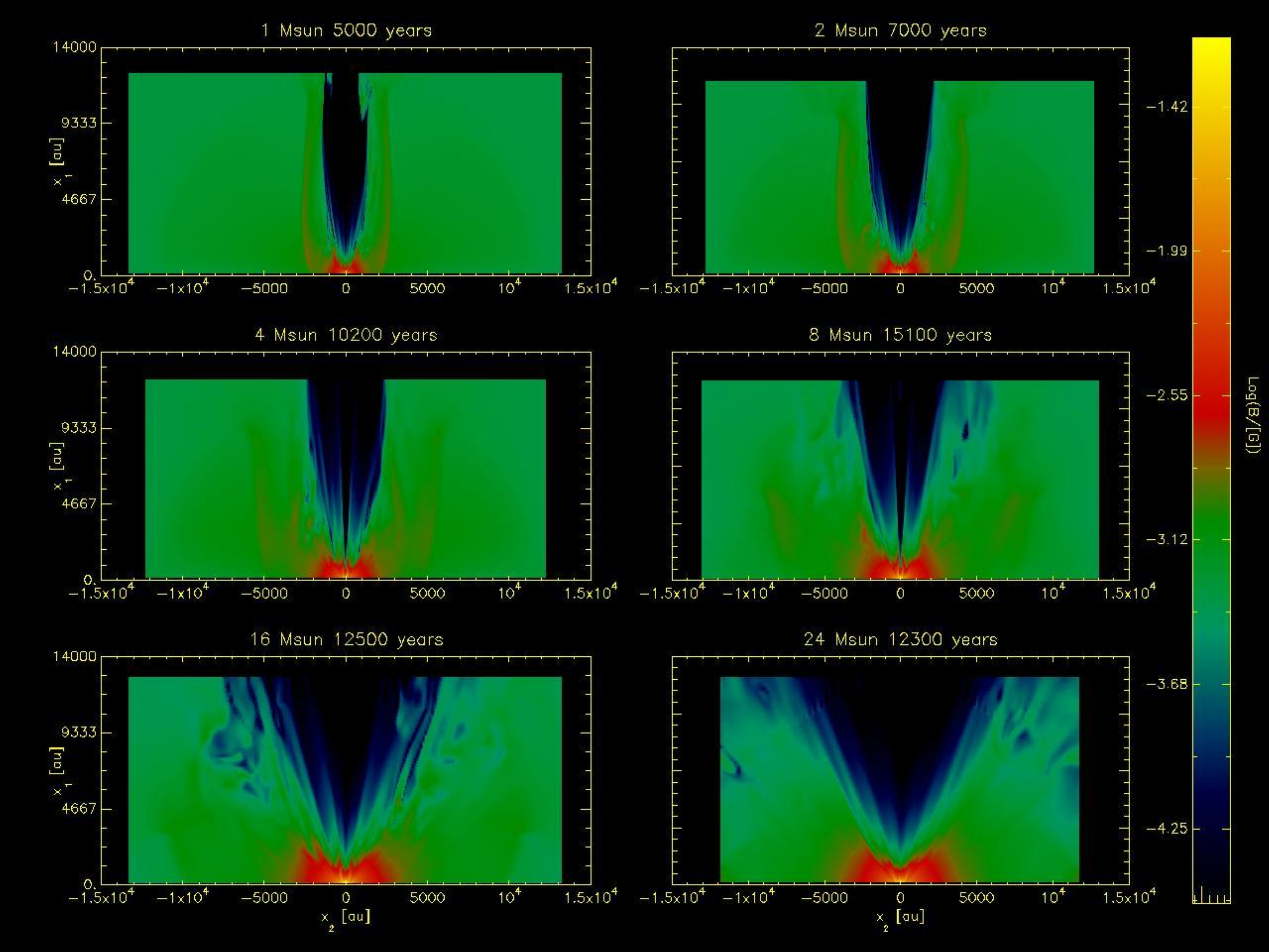}
\caption{
Slices through the middle of the grid showing the logarithm of the
magnetic field strength at the end of each simulation.}
\label{logb}
\end{figure}

In Figure~\ref{logb} we show a slice of the magnetic field
strength at the end of each
simulation.  It is evident that the magnetic field strength within the
outer part of the outflow cavity is relatively weak, with 
$B\ll 1~{\rm mG}$, i.e.,
much lower than the background core's ambient magnetic-field. The
magnetic-field strengths at the base of the outflow are much stronger,
with values approaching $\sim100$~mG in the highest mass cases.

\begin{figure}
\includegraphics[width=0.99\textwidth]{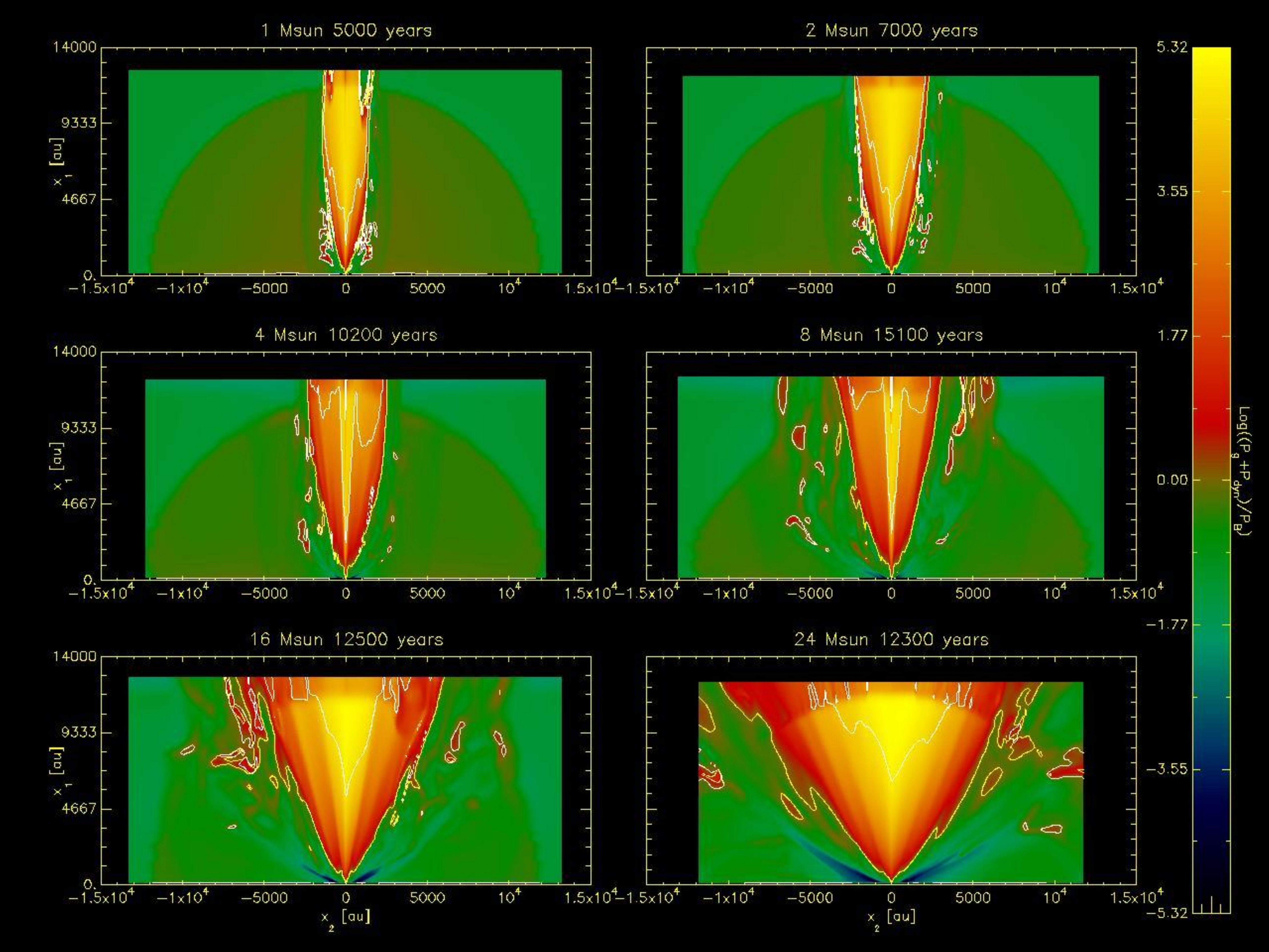}
\caption{Slices through the middle of the grid showing the logarithm of the 
sum of the gas pressure and the dynamical pressure, divided by the magnetic
pressure ($\beta'$), at the end of each simulation. The yellow 
line outlines where the dynamical pressure equals the magnetic pressure. The white line outlines where the gas pressure equals the magnetic pressure.}
\label{gasrambeta}
\end{figure}

In these simulations the plasma $\beta$ ($\beta=P_{\rm gas}/P_{B}$ 
is the ratio of the gas pressure $P_{\rm gas}$ to the magnetic pressure 
$P_{B}$) is for the most 
part much less than unity (i.e., the magnetic pressure dominates).
Dynamical (ram) pressures due to gas flows can be even more important.
Thus, instead of the plasma-$\beta$, we show in Figure~\ref{gasrambeta} the ratio of the sum of the gas pressure and the dynamical pressure to the magnetic pressure, i.e., 
$\beta'=(P_{\rm gas}+P_{\rm dyn})/P_{B}$, where $P_{\rm dyn}=\rho (v_1^2+v_2^2+v_3^2)/2$. 
Figure~\ref{gasrambeta} also shows a 
yellow curve outlining where the dynamical pressure equals the magnetic pressure, and a white curve outlining where the gas pressure equals the magnetic pressure.
In the outflow cavity $\beta'>1$ and the dynamical pressure is by 
far the most dominant.
The exception is for the lowest protostellar masses, where the gas 
pressure is also found to contribute significantly in and around the 
outflow cavity.
Outside of this region, the magnetic pressure is the dominant pressure term
almost everywhere.

\subsection{Outflow opening angle}

\begin{figure}
\includegraphics[width=0.99\textwidth]{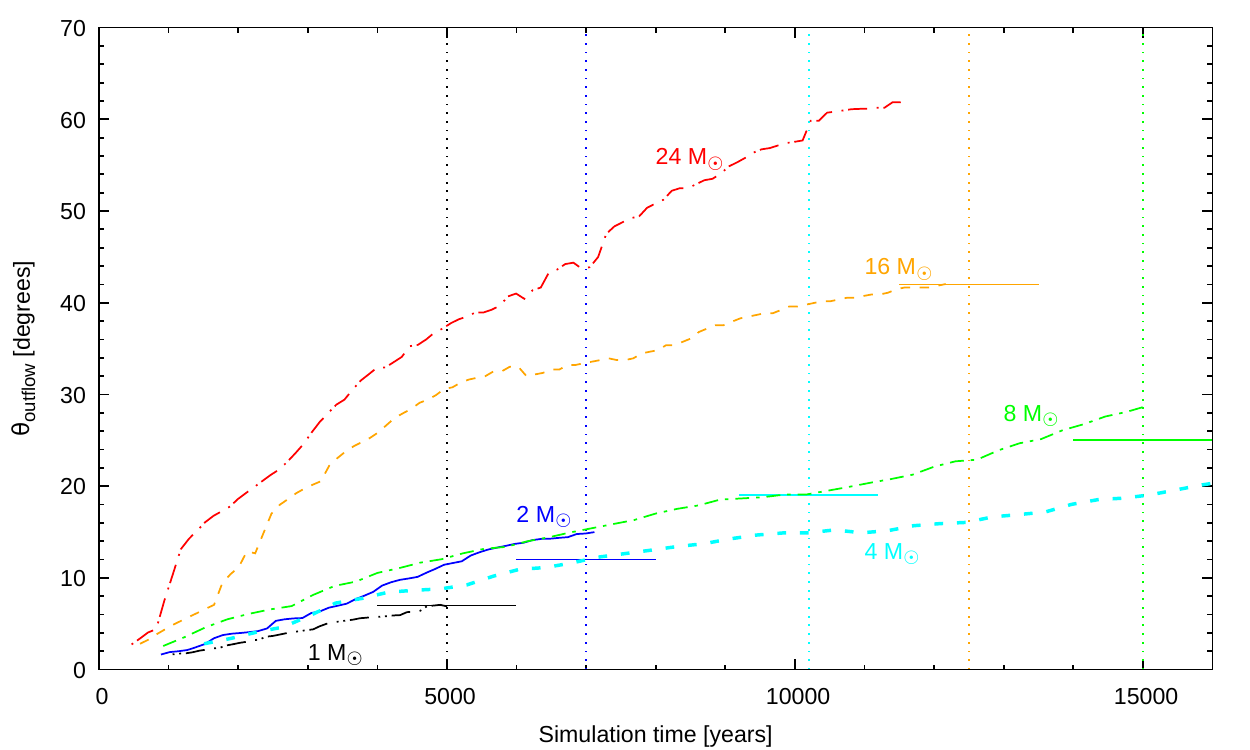}
\caption{
Opening angle of outflow cavity, $\theta_{\rm outflow}$, as a function
of simulation time for the runs with protostellar masses of
$1~{M_\odot}$ (black dot-dot-dash), $2~{M_\odot}$ (solid blue line),
$4~{M_\odot}$ simulation (cyan dashed line), $8~{M_\odot}$ (green
long-dash short-dash line), $16~{M_\odot}$ (orange dashed line) and
$24~{M_\odot}$ (red long-dash dot line) for the case with no
pre-cleared cavity.  The colored vertical dotted lines show the times
when the stars have accreted half the mass needed to take it to the
next simulated model in the sequence.  The horizontal line portions
crossing the vertical lines indicate the opening angles found in the
semi-analytic models of \citet{2014ApJ...788..166Z}.}
\label{openingangletimer}
\end{figure}

\begin{figure}
\includegraphics[width=0.7\textwidth]{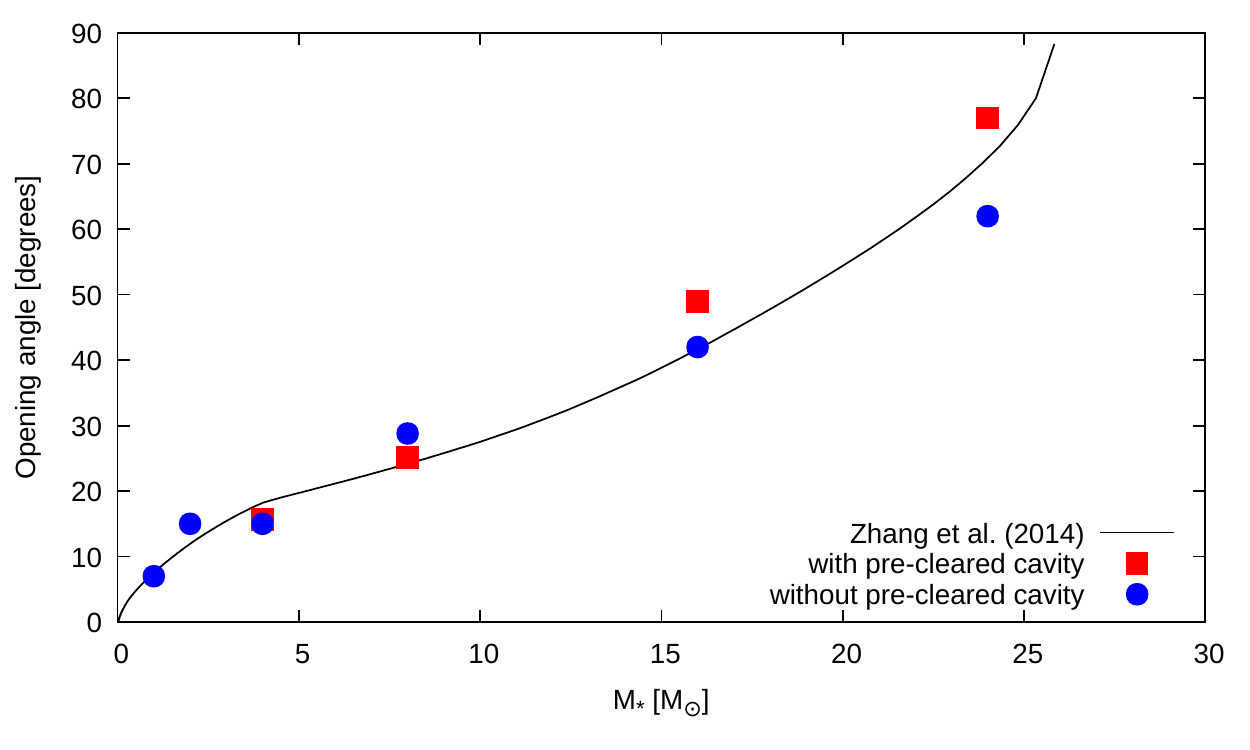}
\caption{
Outflow opening angle versus protostellar mass.  Blue circles are the
opening angles found in our fiducial simulations without
pre-clearing. Red squares are the opening angles in the simulations
with pre-clearing.  The black line shows the results of the
semi-analytic model of \citet{2014ApJ...788..166Z}.  }
\label{openinganglecomp}
\end{figure}

\begin{table}
\centering
\caption{
The final opening angle ($\theta_{\rm outflow}$) found in the
simulations for the various protostellar masses, after a time when the
star would have accreted half the mass needed to take it to the next
simulated protostellar mass (the $24~{M_\odot}$ simulation has
accreted $4~{M_\odot}$).  Also listed is the total mass that has flowed
out of one hemisphere during each simulation.}
\label{openingangletable}
\begin{tabular}{ccccc}
\\
\hline
Mass of star & $\theta_{\rm outflow}$ & $\theta_{\rm outflow}$ & Total mass outflow & Total mass outflow \\
& w/o pre-clearing & w/ pre-clearing 
& w/o pre-clearing & w/ pre-clearing \\
$[{M_\odot}]$ & [degrees] & [degrees] &
$[{M_\odot}]$ & $[{M_\odot}]$ \\
1 & 8.4 & - &  0.35 & -\\ 
2 & 15.0 & - &  0.88 & -\\
4 & 14.9 & 15.7 & 0.70 & 0.28 \\
8 & 28.8 & 25.1 & 1.60 & 1.06 \\ 
16 & 42.0 & 49.0 & 2.75 & 2.05 \\
24 & 62.0 & 77.0 & 3.34 & 2.38 \\
\hline\\
\end{tabular}
\end{table}

Figure~\ref{openingangletimer} shows the time evolution of 
$\theta_{\rm outflow}$ in each of the fiducial simulations, i.e., without
pre-clearing. In each individual simulation $\theta_{\rm outflow}$ grows
from zero from the time when the outflow has just managed to break out
of the initial core envelope structure, typically after $\lesssim
1,000\:$years, depending on $m_*$.  In the case of the $m_*=$1, 2, 4 and
8~$M_\odot$ runs, $\theta_{\rm outflow}$ then increases fairly steadily.
For the 16~$M_\odot$
case, there is more rapid initial expansion as the outflow cavity is
established, and then a more distinct phase of gradual
widening. Finally for the 24~$M_\odot$ case, the expansion is fast and
quite steady for the full duration of the simulated period, with only
a modest decrease in the rate of expansion during the later evolution.

As described above, a natural time to consider the outputs of the
simulation is after the protostar has had sufficient time to increase
its mass significantly, i.e., half-way towards the next model in the
sequence\footnote{Note in the case of the $24~{M_\odot}$ protostar we list 
the result at a time of 12,000~yr, i.e., after it has had time to accrete 
$4~{M_\odot}$.}. These times are used to evaluate the ``final'' 
$\theta_{\rm outflow}$ that is listed in Table~\ref{openingangletable}. In
Figure~\ref{openingangletimer}, these times are marked by vertical dotted
lines. Also shown in this figure are the values of $\theta_{\rm outflow}$ 
expected in the semi-analytic model of
\citet{2014ApJ...788..166Z}. At our adopted output times, these
semi-analytic estimates compare very well with those of our MHD
simulations: they are generally within about 5 degrees of each other.

At a given time, including our chosen ``final'' output times,
$\theta_{\rm outflow}$ is generally larger for models with more
massive protostars. This can also be seen in
Figure~\ref{openinganglecomp}, which shows $\theta_{\rm outflow}$ versus
protostellar mass. The main exception is the $4\:M_\odot$ case, which
has a slightly smaller $\theta_{\rm outflow}$ compared to the
$2\:M_\odot$ case. The reason for this is that at $4\:M_\odot$ the
protostar has evolved into a relatively large, expanded size, due to 
redistribution of entropy in the protostar, including effects of D shell burning \citep{1991ApJ...375..288P,1992ApJ...392..667P,2009ApJ...703.1810H,2010ApJ...721..478H}. This
means the Keplerian speed at the disk inner edge is relatively low so
that the disk wind outflow is relatively weak.  For this case, the
effects of pre-clearing, discussed below, are
expected to be more important.
Figure \ref{openinganglecomp} also compares our simulation results to
the semi-analytic model estimates of \citet{2014ApJ...788..166Z}. It 
is apparent that the numerical results agree well with the semi-analytic 
model predictions.

In addition to the simulations described above, we performed a
sequence of simulations where the initial setup has a pre-cleared
cavity (starting with the $4~{M_\odot}$ simulation) (see
\S\ref{preclearedsection}).  
We find that $\theta_{\rm outflow}$ (see Table~\ref{openingangletable} and Figure~\ref{openinganglecomp}) is not much different whether or
not we have pre-cleared a cavity, which indicates the general
robustness of the results and the validity of ignoring prior evolution
for each fiducial simulation run.

However, $\theta_{\rm outflow}$ in the $24~{M_\odot}$ simulation with
pre-clearing does open up faster and to a moderately greater extent ($\theta_{\rm
  outflow}=77.0^\circ$) than in the simulation without
pre-clearing ($\theta_{\rm outflow}=62.0^\circ$).
Interestingly, in the $8~{M_\odot}$ simulation, $\theta_{\rm outflow}$
is smaller with pre-clearing than without. This may be due to the
outflow feedback being more easily directed into the low-density
initial cavity, i.e., deflecting off the dense core infall envelope,
and so more easily confined. In the case without the initial cavity,
the outflow may be able to establish a broader opening angle during
its initial break-out phase.

\subsection{Mass flow rate and momentum rate}

\begin{figure}
\includegraphics[width=0.47\textwidth]{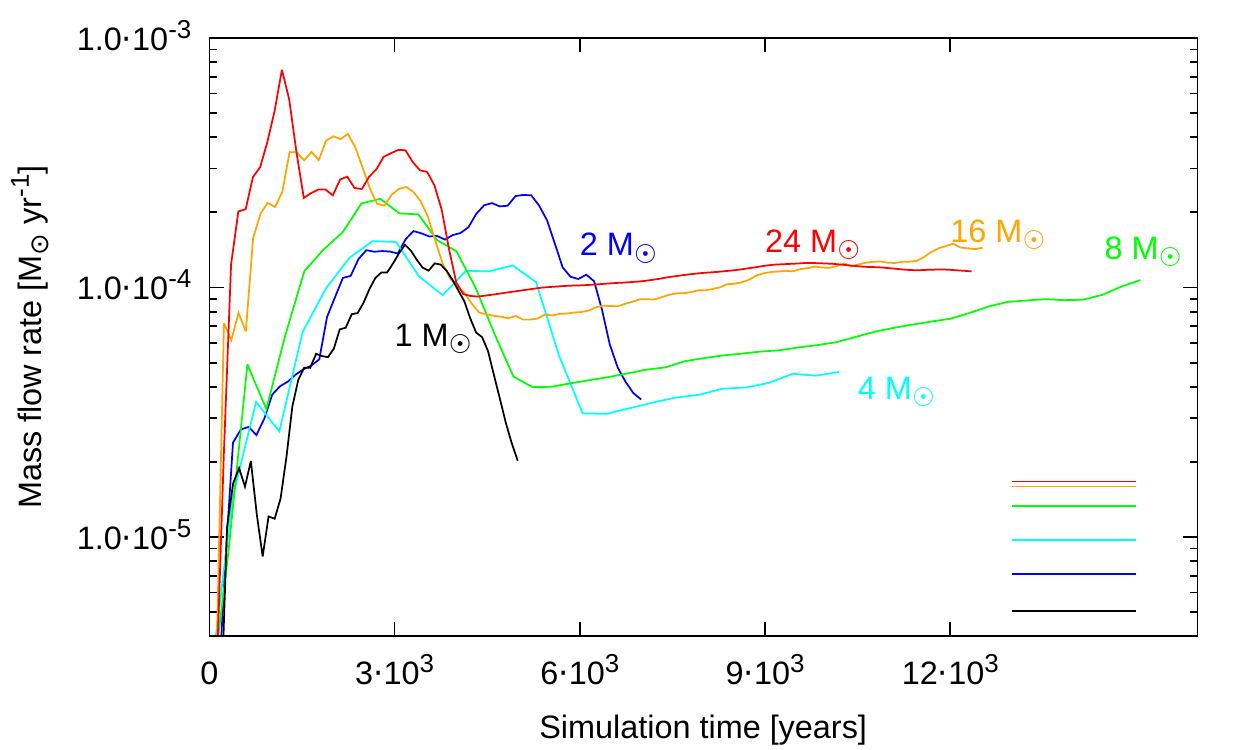}
\includegraphics[width=0.47\textwidth]{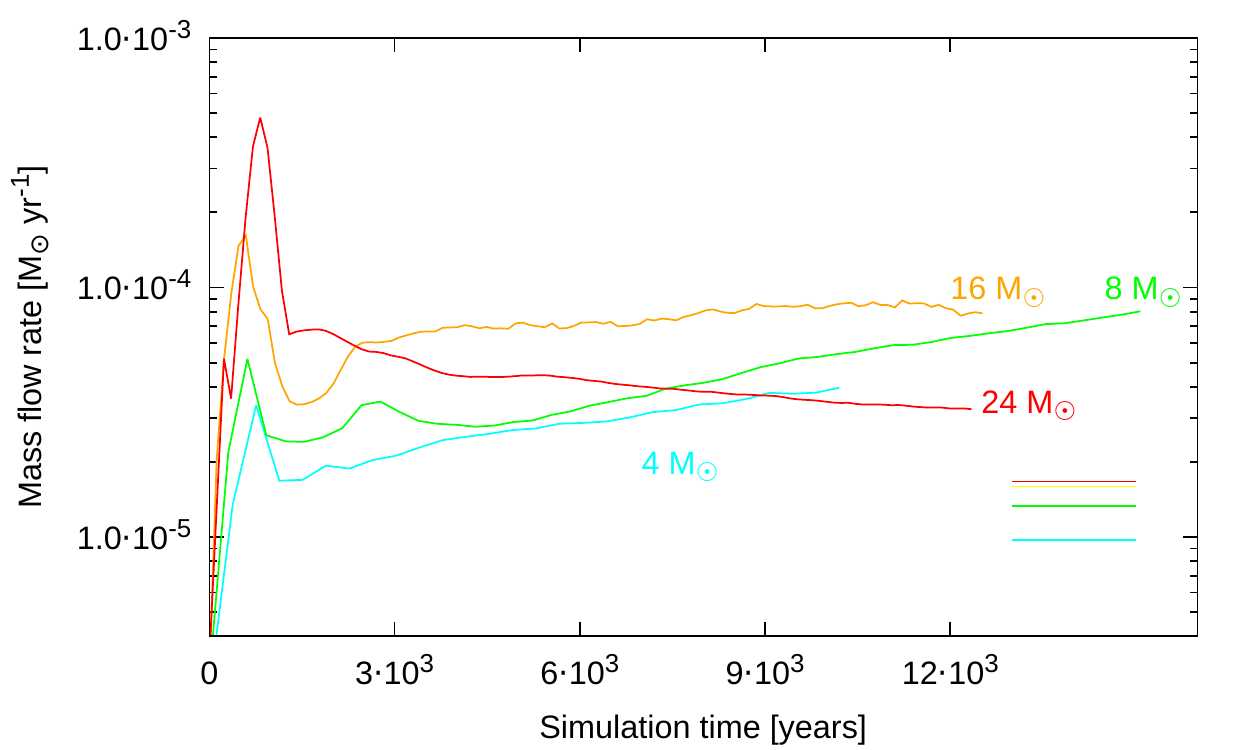}
\caption{
The mass flow rate (in one hemisphere) as a function of time for the 
simulations without
pre-clearing (left panel), and with pre-clearing (right panel). 
The horizontal lines in the right side of the figures show the
injected mass flow rate, for comparison.}
\label{mfluxtime}
\end{figure}

\begin{figure}
\includegraphics[width=0.47\textwidth]{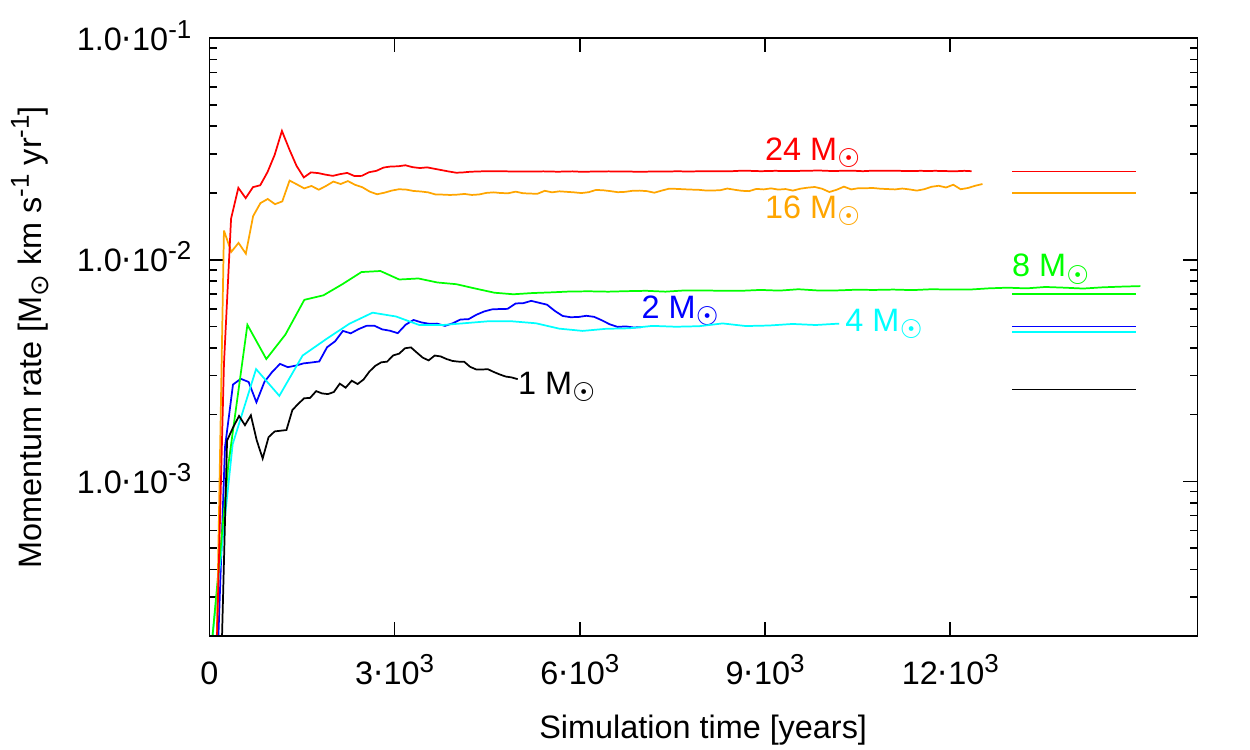}
\includegraphics[width=0.47\textwidth]{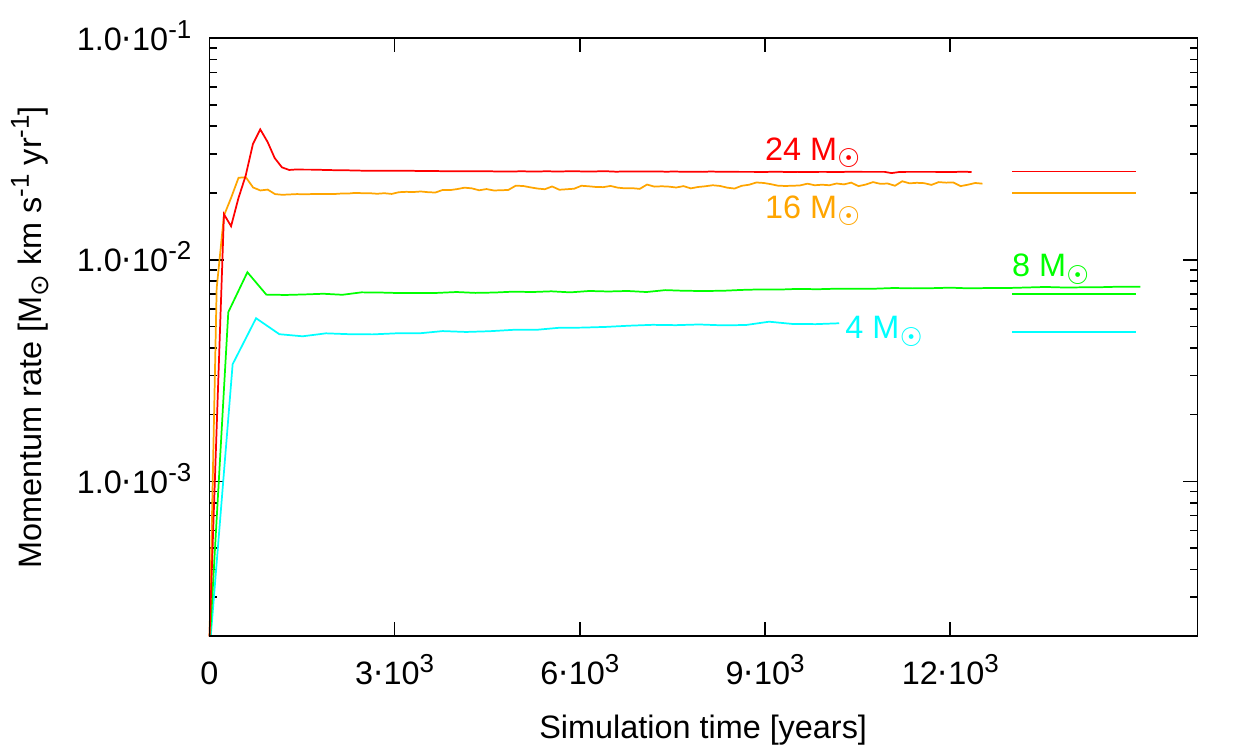}
\caption{
The momentum flow rate (in one hemisphere) as a function of time for 
the simulations without
pre-clearing (left panel), and with pre-clearing (right panel).
The horizontal lines in the right side of the figures show the injected
momentum rate, for comparison.}
\label{momfluxtime}
\end{figure}

\begin{figure}
\includegraphics[width=0.98\textwidth]{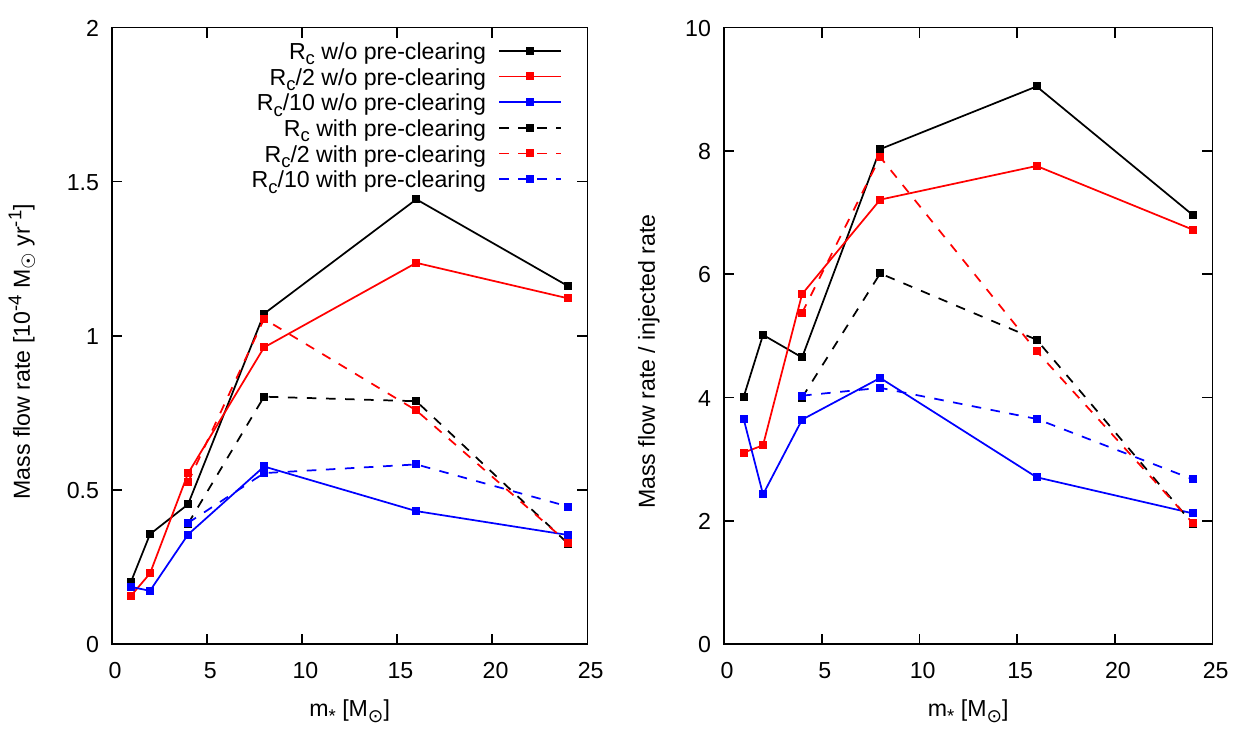}
\caption{
Mass outflow rate at end of the simulations (in one hemisphere) at a height above the disk of 
$1/10~R_{\rm c}$ (blue
curve), $1/2~R_{\rm c}$ (red curve), and $R_{\rm c}$
(black curve) as a function of the protostellar mass (left
panel). The right panel show the same, but normalized to the
injected flow rate (into one hemisphere). The solid lines are the 
simulations without a
pre-cleared cavity, while the dashed lines are  the simulations with a
pre-cleared cavity. }
\label{massflux}
\end{figure}

\begin{figure}
\includegraphics[width=0.98\textwidth]{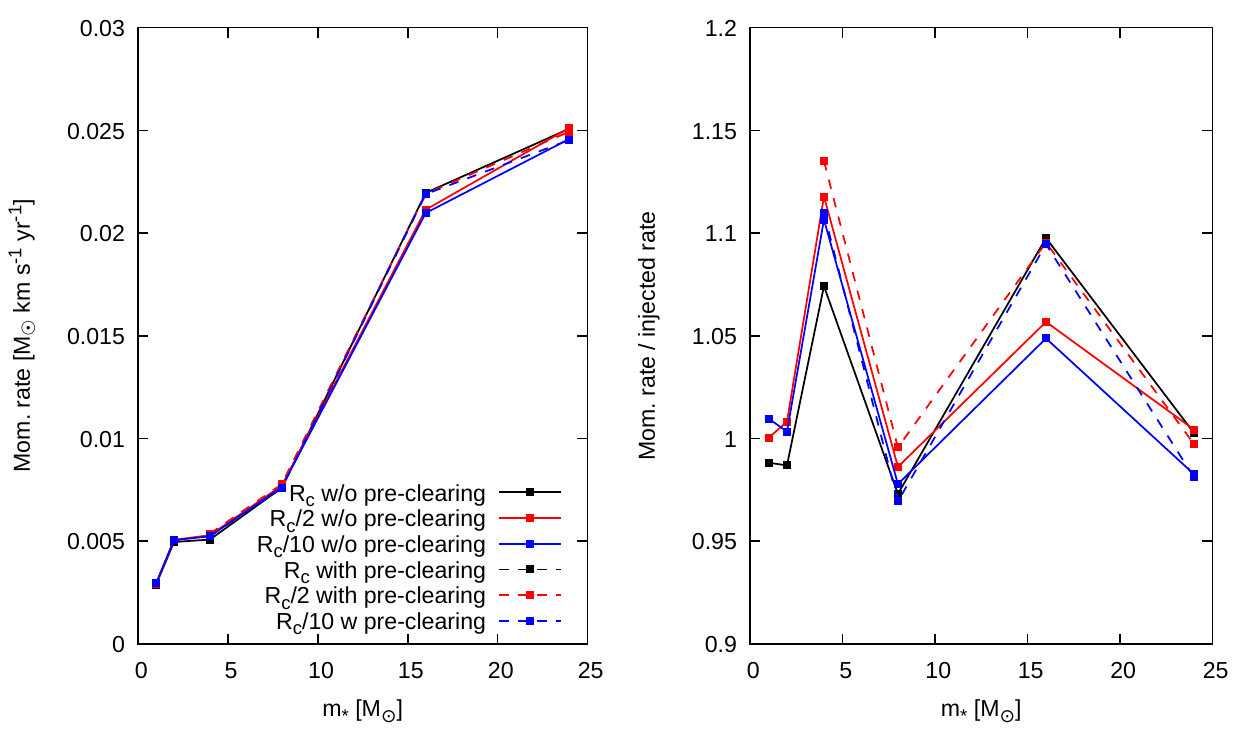}
\caption{
The outflowing momentum rate at the end of the simulations (in one hemisphere) at a height above the disk 
of $1/10~R_{\rm c}$ (blue
curve), $1/2~R_{\rm c}$ (red curve), and $R_{\rm c}$
(black curve) as a function of the protostellar mass (left
panel). The right panel show the same, but normalized to the
injected rate (into one hemisphere). The solid lines are the simulations without a
pre-cleared cavity, while the dashed lines are the simulations with a
pre-cleared cavity.  }
\label{momflux}
\end{figure}

Figures~\ref{mfluxtime} and \ref{momfluxtime} show the mass outflow
and momentum flow rates measured at a height of $R_{c}$ above the disk,
as a function of time for our
simulations.  
Here the mass outflow rate ($\dot{m}=\int\rho v dx_2 dx_3$, for $v_1>c_s$) is the mass flowing out of one hemisphere in the $x_1$ direction only, and likewise the momentum flow rate ($\dot{p}=\int\rho v^2 dx_2 dx_3$, for $v_1>c_s$) is only including the momentum in the $x_1$ direction, again from one hemisphere.
Without pre-clearing, the flow rate for
each simulation has a large ``bump'' in the early part of the
simulation.  
This bump is a result of the initial state. In each simulation, the flow has
to clear out a new outflow channel, leading to this artificial transient 
event. Once an outflow channel has been established by pushing the 
mass out of the simulation box, the flow rate stabilizes. 
The effect of the pre-clearing is apparent in the 
mass outflow rate figure.  
With pre-clearing, there is still a ``bump'', but it is
much less prominent.  
We see from Figure~\ref{mfluxtime} that 
the mass flow rate out of the core at the 
end of the simulations is always larger than
the injected mass flow rate by a factor of a few.
The larger flow rate out of the core is due to erosion of the infalling 
envelope.
In the $24~{\rm M_\odot}$ simulation with pre-clearing, the mass flow rate 
gradually drops. This is due to there not being much mass to sweep up.

The momentum flow rates also
stabilize after an initial transient phase
($\sim5-10\times10^{-3}~M_\odot~{\rm km~s^{-1}\:yr^{-1}}$
for the 4 and 8 ${M_\odot}$ simulations and
$\sim2\times10^{-2}~M_\odot~{\rm km~s^{-1}\:yr^{-1}}$ for the 16 and
24 ${M_\odot}$ simulations).  
The effect of
the pre-clearing is also visible in the momentum rates, in that these
are smoother at earlier times with pre-clearing than without. 
The mass flow and momentum rates stabilizes around the same values in the
simulations with and without pre-clearing.
This value is roughly equal to the injected momentum rate.

In Figure~\ref{massflux} we show the mass outflow rate at the end of each 
simulation at different
heights above the disk in the envelope (at $R_{c}$, $R_{c}/2$, and $R_{c}/10$) 
as a function of the protostellar mass for both the
simulations without and with pre-clearing. 
We find that the mass
outflow rate at $x_1=R_{c}$ generally is larger for larger protostellar 
mass.
This is mainly due to the stronger injection (see Table~\ref{setupparamtable}).
As the outflowing material makes its way through the envelope, mass is
being swept up. As a consequence, the mass flow rate at $R_{\rm c}$ is
a factor 4-10 times the injected mass flow rate, it is
generally larger than deeper in the core, and it is larger in the simulations
without pre-clearing.

Figure~\ref{momflux} shows the outflow momentum rate at the end of each 
simulation at different
heights above the disk in the core as a function of protostellar mass.  
We find that
the momentum flow rate increases as $m_*$ rises, reaching a few $\times 10^{-2}~M_\odot~{\rm
  km~s^{-1}\:yr^{-1}}$, and that it remains approximately constant as the flow
propagates through the core, starting from the injection boundary. 

The total mass flowing out of one hemisphere (found by
integrating the mass outflow rate over time)
for each case is listed in Table~\ref{openingangletable}.
Summing all the simulations without pre-clearing, we find that
$9.62~{M_\odot}$ flowed out of one hemisphere in our simulations.  With
pre-clearing, the sum is $5.77~{M_\odot}$, though that excludes the
$1$ and $2~{M_\odot}$ simulations.  Since the simulations only run to
half-way to accreting to the next stage, then the total mass
ejected if continuous growth of the protostar were followed is
expected to be about twice the above values, i.e., $2\times
\sim6\:M_\odot \simeq 12\:M_\odot$ in the case with
pre-clearing. Accounting for both hemispheres, the total outflowing mass becomes $\sim24~{M_\odot}$, which is similar to the
mass growth of the protostar, demonstrating that the star
formation efficiency is about 50\% from the core during this
evolution.

\subsection{Effects of injected flow rotation and an unmagnetized outflow}

\begin{figure}
\includegraphics[width=0.9\textwidth]{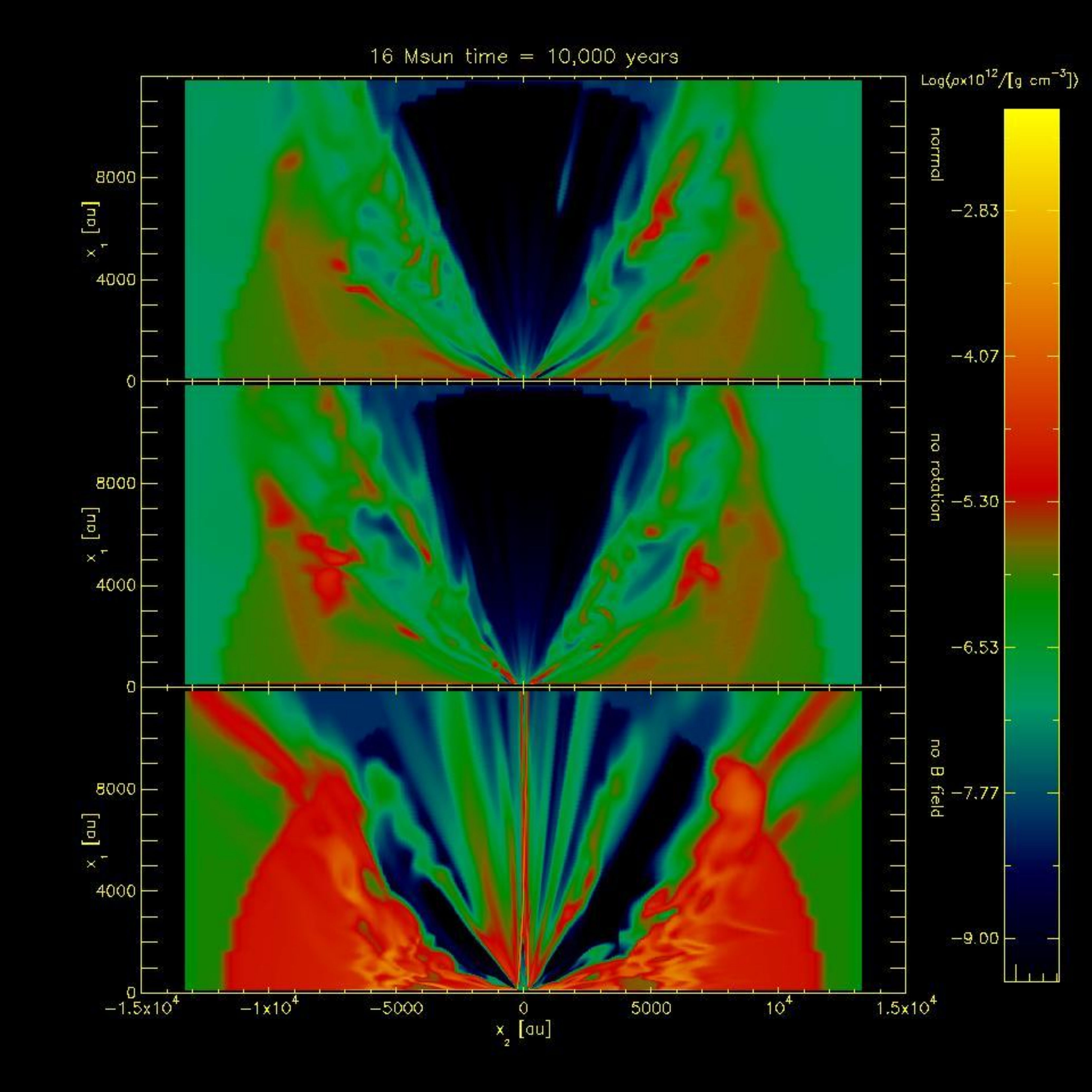}
\caption{
The $16~{M_\odot}$ simulation after $10^4$ years. The top panel
shows the normal simulation, the middle panel shows the same setup
except that the injected material is not given a rotational
velocity. The bottom panel shows the same setup as the top panel, but
with no magnetic field.}
\label{comp16sim}
\end{figure}

We find that in the $16~{M_\odot}$ simulation (without pre-clearing),
after $10^4$ years, there is very little difference whether the
injected material is given rotation or not (see Figure~\ref{comp16sim}).
In both cases the opening angle at this particular time is
$42^\circ$.  

In a test simulation of the same $16~{M_\odot}$ case without magnetic field we find that the opening
angle is $48^\circ$.  
This illustrates the role of the magnetic pressure in confining the flow. 
Without a magnetic field, the flow will open up until $P_{\rm dyn}$ is approximately
balanced by $P_{\rm gas}$.
Accordingly, we also find that
the no magnetic field simulation has a larger outflowing mass rate
than the simulation with magnetic field, due to the larger opening
angle of the outflow.  The
simulation with magnetic field has a mass flow rate out of one hemisphere of
$2.2\times10^{-4}~{M_\odot{\rm\:yr^{-1}}}$, while that without 
magnetic field has a mass flow rate of $3.1\times10^{-4}~{M_\odot{\rm\:yr^{-1}}}$, i.e., 1.4 times greater. 


One noticeable difference, however, 
is that without a magnetic field, the outflow is not capable
of maintaining a clear outflow cavity.
Another difference is that the density of the remaining envelope 
material outside of the outflow cavity is an order of magnitude larger in
the simulation without magnetic field, as seen in Figure~\ref{comp16sim}. 
Without the magnetic field to confine and collimate the outflow,
the wider angle flow interacts with more of the collapsing envelope,
pushing and compressing it and thus causing the higher densities seen in the figure.

\subsection{Dependence on numerical resolution}
\label{ressubsection}
\begin{figure}
\includegraphics[width=0.9\textwidth]{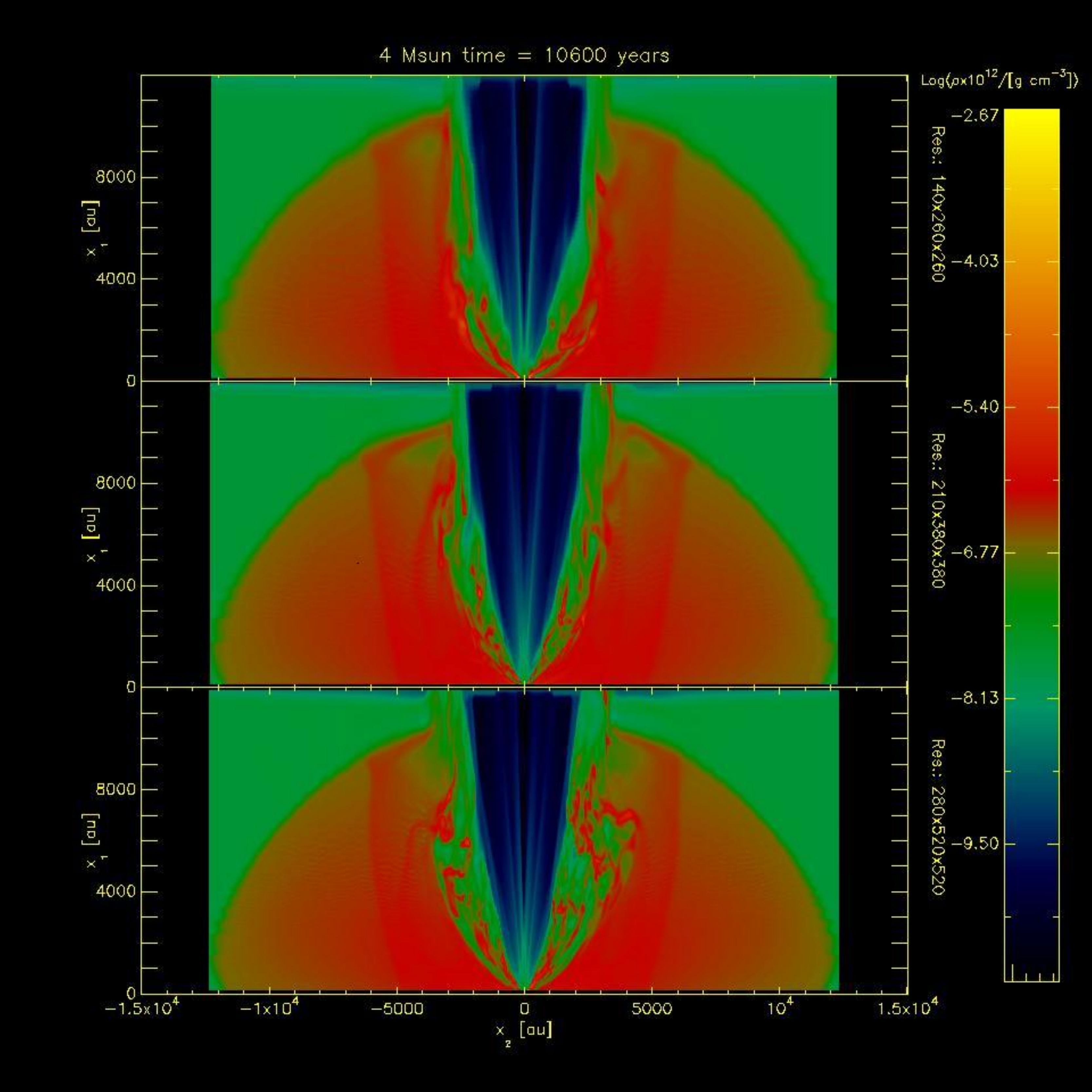}
\caption{
Comparison of the density structure through the middle of the grid of
the $4~{M_\odot}$ simulation in normal resolution (top panel),
medium resolution (middle panel), and high resolution (bottom panel)
after 10,200 years of simulation time. }
\label{4Mres}
\end{figure}

As described in \S\ref{methodssection}, we used a grid with 
$140\times260\times260$ cells in our standard
setup for these simulations (with $m_*\geq4~{M_\odot}$).
To test the effect of grid resolution on the results, we
also ran the $4~{M_\odot}$ simulation using a grid with
$210\times380\times380$ cells (medium resolution), and using a grid
with $280\times520\times520$ cells (high resolution).
Figure~\ref{4Mres} compares the results of these
simulations after 10,200 years of simulation time, at which point the
protostar would have accreted $2~{M_\odot}$.  While some differences
are apparent in the density structures, the opening angle of the flow is found
to be approximately the same in all three cases.  In particular,
$\theta_{\rm outflow}$ in the standard, medium and high resolution simulations was found to
be $14.9^\circ$, $16.5^\circ$ and $18.2^\circ$, respectively.


\begin{figure}
\includegraphics[width=\textwidth]{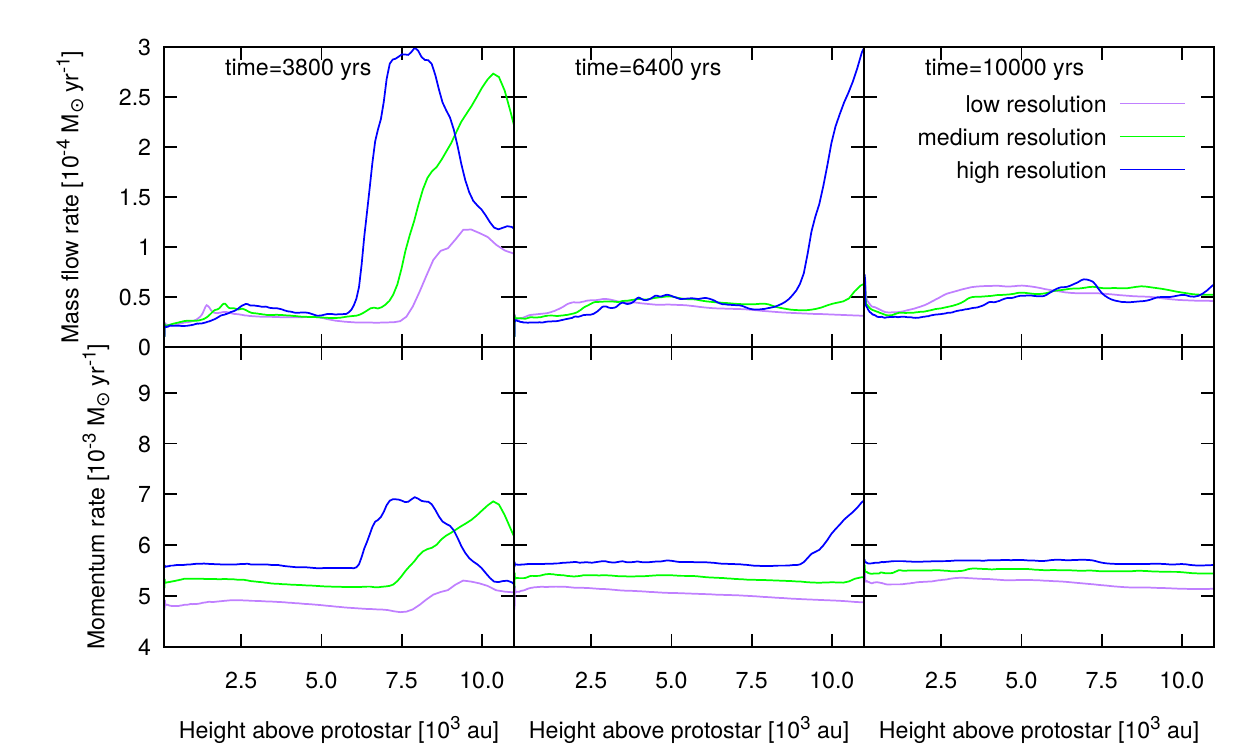}
\caption{The mass flow rate (top row), and momentum rate (bottom row) 
throughout the core at three different times in the
$4~{\rm M_\odot}$ simulation, for the low resolution simulation (purple),
medium resolution simulation (green), and high resolution simulation (blue).}
\label{massfluxseq}
\end{figure}

In all of these resolution test runs, the flow begins to break out of the core after
approximately 1,000 years.  We find that the mass
flow rate, differs a lot between
the different resolutions at earlier times 
(see Figure~\ref{massfluxseq}).  
After 6,400 years we find
the largest difference, with the mass flow rate out of the core in the 
high resolution
simulation being almost an order of magnitude larger than in the low-resolution
simulation.  However, after 10,000 years the mass flow rates in the different
resolution simulations appear to be converging towards
$\approx5\times10^{-5}~{M_\odot{\rm\:yr^{-1}}}$.  
The difference is related to the clearing of the outflow cavity, with
the higher resolution run taking longer time to push the material out
of the simulation box.

The momentum rate does not show the same level of
differences, and after 6,400 years the momentum rate in the high
resolution simulation is only a factor 1.5 larger than in the low-resolution 
simulation, which
is the largest deviation between the simulations that we find.
After 10,000 years the momentum rate in the high resolution simulation
is $\approx10\%$ larger than in the low resolution simulation.
Again the differences are related to the clearing of the outflow 
cavity, see Figure~\ref{massfluxseq}. The slightly higher momentum rates in
the high resolution simulation are due to the larger velocities being
resolved in the central regions.
We are
therefore satisfied that the resolution does not significantly affect
the results of our simulations.
We caution that neither the mass flow rate nor the momentum rate is
constant throughout the grid, so these values are subject to the exact 
slice and time at which they are calculated.  

We recall that we use a logarithmic grid, and that we aim at resolving 
the injection radius $r_{\rm inj}$ with 10 cells (see \S\ref{methodssection}).
Since in the $1$ and $2~{M_\odot}$ simulations the
injection region is the smallest, these simulations therefore have the
smallest cells (in the $x_2$ and $x_3$ directions), which therefore 
necessitates the biggest stretching of the grid in order to extend the 
grid beyond $R_{\rm c}$, when using the same 
number of cells.  
The cells around the axis near the inner $x_1$ boundary are cubic 
by design, so this also lead to small cells in the $x_1$ direction 
near the injection boundary.
In these $1$ and
$2~{M_\odot}$ simulations, we found that when using the same resolution as
the higher protostellar masses, the envelope develops a
``noisy'' density structure over time.  This turned out to be related
to the degree of stretching of the grid that we had to utilize to both 
resolve the injection radius with $\sim10$ cells, and resolve the whole 
envelope structure.  
Using more grid cells
(in all directions, but in particular in the $x_1$ direction), we found 
that we could reduce
the noisiness of these simulations, however, the main results, i.e., $\theta_{\rm outflow}$ and $\dot{m}$, were not
affected.  

\section{Discussion and Summary}

In our presented MHD simulations, we follow a sequence of evolutionary
models of the protostar, enabled by injecting a disk wind into the
simulation box at a height of $100~{\rm au}$ above the disk.
A number of other groups have
performed collapse simulations to study outflows from massive
protostars, using various numerical techniques and including different
physics.  However, most of these do not follow the evolution of the
protostar until the end, and therefore can not estimate the full evolution of the morphology, outflow properties and the star
formation efficiency.

\subsection{Comparison to previous theoretical work}

Some previous studies performed ideal-MHD simulations of $100~{M_\odot}$ core collapse \citep{2011MNRAS.417.1054S,2012MNRAS.422..347S,2011A&A...528A..72H,2011ApJ...742L...9C}. Those MHD simulations showed especially that fragmentation is suppressed by magnetic pressure and magnetic breaking in the highly magnetized cases. However, they could not continue the simulation long enough to reach a $\sim 5~{M_\odot}$ protostellar mass due to the high numerical cost of following the small-scale processes, i.e., disk formation and outflow launching.

The collapse of a massive cloud core was also simulated by
\citet{2017MNRAS.470.1026M,2018MNRAS.475..391M}. Starting from a range of 
cloud masses, they followed the
protostar until it reached a mass of $\lesssim30~{\rm M_\odot}$.
For their simulations with cloud masses of $32$ and $77~{\rm M_\odot}$ 
(their simulations most comparable to our setup),
they stopped the simulations when the protostar was a few solar masses.
At their highest resolution, they have a
resolution of $0.8~{\rm au}$, a factor of 6 smaller
than in our highest resolution simulations.
In their simulation with a cloud mass of $77~{M_\odot}$, they found a similar mass ejection rate ($\sim10^{-4}~{M_\odot{\rm\:yr^{-1}}}$)
to that which we found in our models.

However, since these works terminate the calculations before the 
star reaches its 
final mass, they do not estimate the star formation efficiency.
This is one of the main objectives of our paper. We therefore
simulated a sequence of models using boundary conditions relevant to the 
Turbulent Core Model that can be compared to the semi-analytic work of 
\citet{2014ApJ...788..166Z}.

We have found that the opening angle increases with more massive
protostars, i.e., with age, in agreement with the
evolutionary sequence proposed in \citet{2005ASSL..324..105B}.  
Building on the model of \citet{2000ApJ...545..364M},
\citet{2014ApJ...788..166Z} evaluated the evolution of the outflow
opening angle during the growth of a massive protostar.  For their
fiducial values, they found that a core with initial mass of
$60~{M_\odot}$ reaches a stage with an $8~{M_\odot}$ protostar with an
outflow with opening angle of $25^\circ$. Later it grows to a
$16~{M_\odot}$ protostar having an outflow with an opening angle of
$40^\circ$.  The final star resulting from their model had a mass of
$26~{M_\odot}$.  This is in reasonably good agreement with what we
find in our study.
We note that this is also in agreement with observations 
\citep{2007prpl.conf..245A}. 

Since we find similar results to those in \citet{2014ApJ...788..166Z}
for outflow opening angles, we therefore also obtain a SFE of $\sim
50\%$, similar to what they found.  This is then an indication that such MHD
disk winds may be a dominant mechanism for limiting the
growth of the protostar and ultimately helping to shape the stellar
initial mass function from a given pre-stellar core mass function.

Radiative feedback is also expected to have a significant impact on the formation of massive stars \citep{2009Sci...323..754K,2010ApJ...722.1556K,2016ApJ...823...28K,2016MNRAS.463.2553R}.  However, the magnetically-driven outflow creates the cavity before the luminosity becomes sufficiently high to interfere with the mass accretion. Since the outflow cavity channels the radiation, the impact of radiative feedback is reduced \citep{1999ApJ...525..330Y,2005ApJ...618L..33K,2015ApJ...800...86K}. Recent studies including multiple feedback processes together show, at least in the case of cloud cores with $M_c<100~{M_\odot}$, that the MHD disk wind is likely to be the dominant feedback mechanism determining the SFE \citep{2017ApJ...835...32T,2018arXiv180410211K}. Based on these results, we conclude that the radiative processes would not alter our results significantly.

We note that a limitation of applying our work to link CMF and IMF is
that we have focused on single stars, but most massive stars are in binaries \citep{2012Sci...337..444S}.
\citet{2017MNRAS.470.1626K} simulated outflows from the formation of
both single and binary stars.  While their simulations could not
predict the final masses of the stars due to computational
limitations, they found that the single star case accreted less mass
compared to the binaries in the same amount of time.  Hence single
stars may have a lower star formation efficiency than binary stars.
Such an effect may be due to the relatively weaker outflows from two
lower mass protostars compared to that of a single protostar with
twice the mass.

\subsection{Comparison to observations}

\begin{figure}
\includegraphics[width=\textwidth]{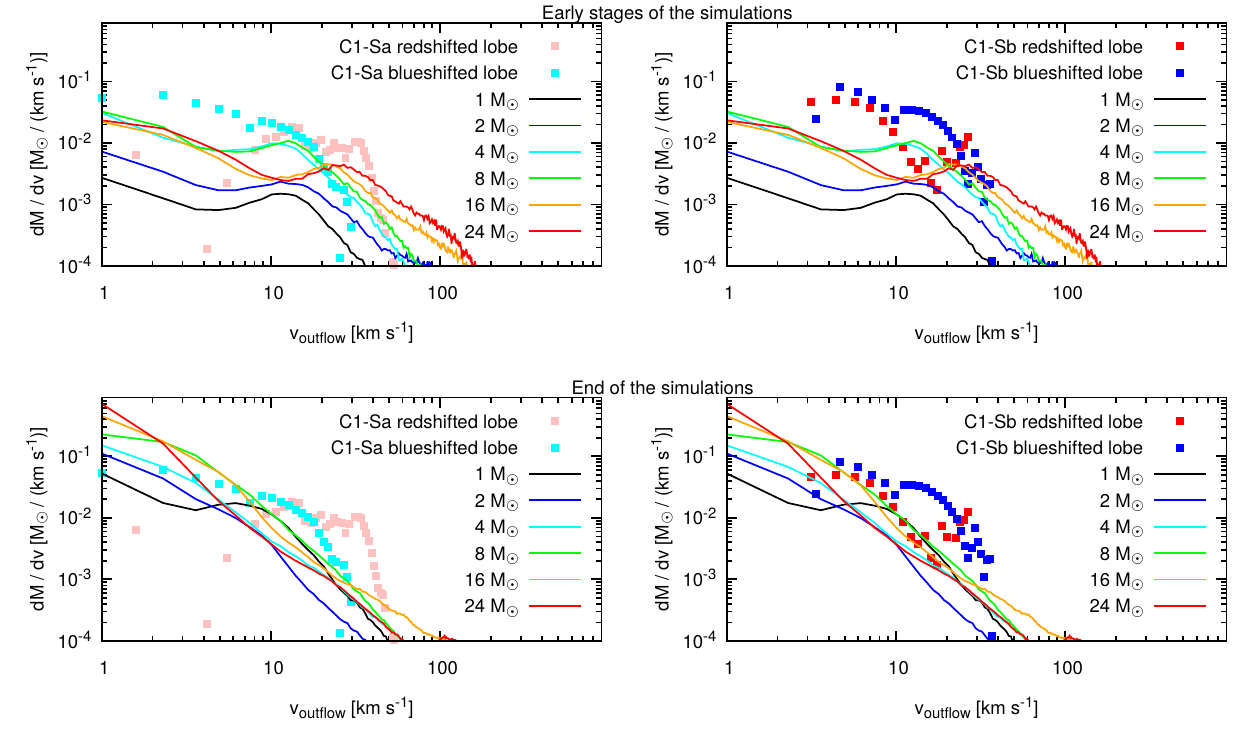}
\caption{The same as Figure~\ref{histograms}; histograms showing the 
distribution of the outflow mass with respect to the outflow speed, 
evaluated around the time that the outflow breaks out of the core in 
each simulation (top row), and at the end of each simulation (bottom row).  
In the left panels, observed data of protostar C1-Sa is shown, and in the right panels,
observed data of protostar C1-Sb is shown \citep{2016ApJ...821L...3T}.}
\label{histogramsobs}
\end{figure}

\begin{figure}
\includegraphics[width=0.98\textwidth]{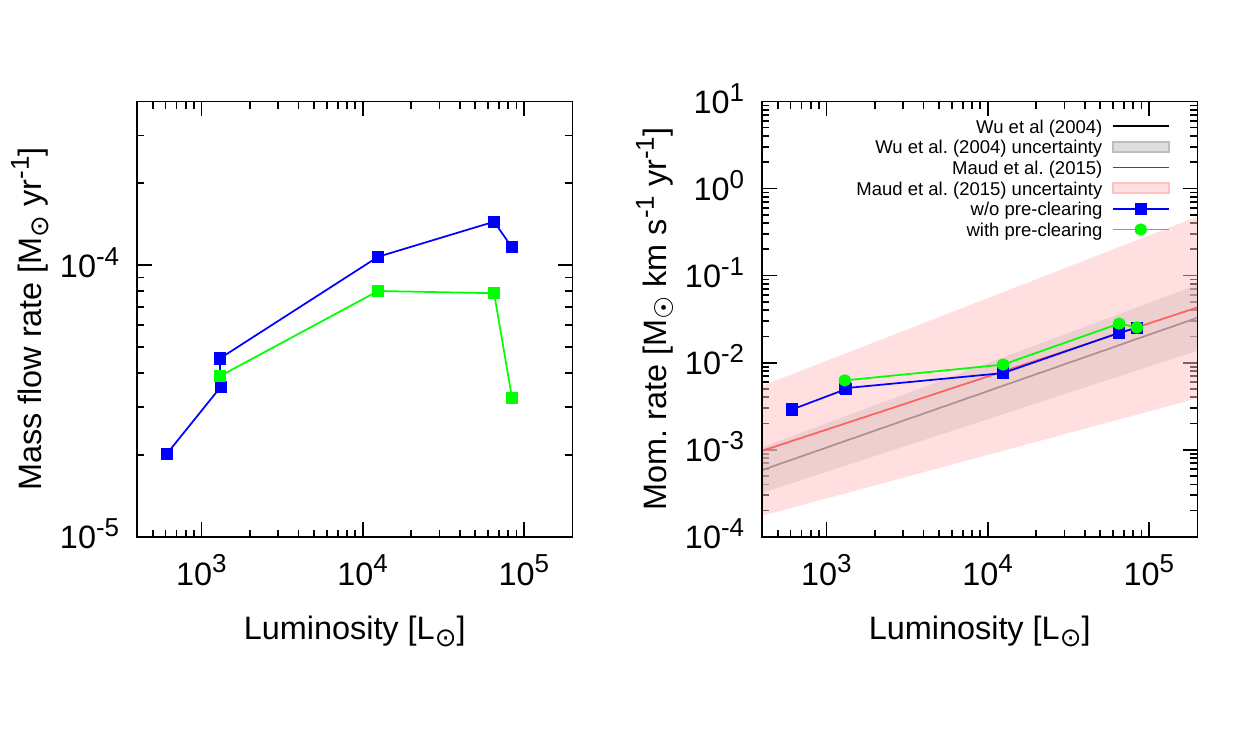}
\caption{
The mass flow rate (left panel) and momentum rate (right panel) in
one hemisphere as a
function of bolometric luminosity. The blue curve is for the
simulations without pre-clearing, the green curve is for the
simulations with pre-clearing. The solid gray line is the best fit from
\citet{2004A&A...426..503W}, the gray shaded area indicates the uncertainty,
while the solid red line is the best fit from
\citet{2015MNRAS.453..645M} and the pink shaded area indicates the uncertainty. Note that there are two blue points roughly on top of each other in the 
momentum rate plot at a luminosity of $\sim1.3\times10^3~{\rm L_\odot}$}.
\label{dotlbol}
\end{figure}

We have compared the distributions of outflowing masses in our
simulations with some observed cases (Figure~\ref{histogramsobs}), i.e., 
those found in the protostars C1-Sa and
C1-Sb \citep{2016ApJ...821L...3T}. In the blueshifted outflow, they
found the majority of the outflowing material at velocities well below
$10~{\rm km~s^{-1}}$, while in the redshifted outflow it is more
evenly distributed to $\sim30~{\rm km~s^{-1}}$, especially in
C1-Sa. 
The observed distributions 
appear to drop off
with a steeper power law at high velocities than in our simulations.
We find that generally the power law is steeper in the lower
protostellar mass simulations than in the higher mass cases.

C1-Sa's
redshifted outflow and C1-Sb's blueshifted outflow show significantly
larger amounts of mass at higher velocities (and note that these
distributions are not corrected for inclination, so once corrected would actually be even
higher).
We note, however, that the outflow properties from these sources were measured 
on scales extending $12^{\prime \prime}$ from the protostar, or 60,000 au. This is much larger
than our simulation box of $\sim12,000~{\rm au}$. In our simulations, 
much material, and in particular high velocity material, has left the 
simulation box (see Table~\ref{openingangletable}).
We therefore find it interesting that we roughly recover the bumps 
seen in the observed redshifted curves around $10$ and $30~{\rm km\:s^{-1}}$
at early times, before (especially high velocity) mass has been lost from the 
simulation box. These bumps, however, are related to the initial clearing
of the outflow cavity for each simulation in the sequence, since these
bumps are much less prominent in the simulations with pre-clearing. At later 
times, we recover the low velocity components reasonably well.

\citet{2015MNRAS.453..645M} found outflow momentum rates of
$10^{-3}-10^{-1}~M_\odot~{\rm km~s^{-1}{\rm\:yr^{-1}}}$ for 
outflows from core masses of $\sim 60~{M_\odot}$ (luminosity of
$\sim10^3-10^5~{L_\odot}$), while 
\citet{2002A&A...383..892B} also found momentum rates of $\sim
10^{-3}-10^{-1}~{M_\odot~{\rm km~s^{-1}\:yr^{-1}}}$ for such outflows.
This is in reasonable agreement with our results, which reflect our
choice of input boundary conditions. We found momentum rates (leaving
the core) of $5\times10^{-3}-3\times10^{-2}~M_\odot~{\rm
  km~s^{-1}{\rm\:yr^{-1}}}$.  These works also found mass flow rates of the
order $10^{-4}-10^{-3}~{M_\odot~{\rm\:yr^{-1}}}$ for sources with
luminosity of $\sim10^3-10^5~{L_\odot}$, which are also in reasonable
agreement with our mass flow rates found to be $0.5-2.5
\times10^{-4}~M_\odot~{\rm\:yr^{-1}}$.

We show the stabilized (final) mass
flow rates and momentum flow rates as a function of bolometric luminosity
in Figure~\ref{dotlbol}.  The bolometric luminosity has been estimated
based on the mass of the protostar following
\citet{2018ApJ...853...18Z}.  The figure also shows the best fit line
for the momentum flow rate from the observational data in \citet{2004A&A...426..503W},
given by $\log (\dot{P}/({M_\odot~{\rm km~s^{-1}\:yr^{-1}}})) = (-4.92\pm0.15)+(0.648\pm0.043) \log (L_{\rm bol}/{L_\odot})$, and in \citet{2015MNRAS.453..645M} given by 
$\log (\dot{P}/({M_\odot~{\rm km~s^{-1}\:yr^{-1}}})) = (-4.60\pm0.46)+(0.61\pm0.11) \log (L_{\rm bol}/{L_\odot})$.

We found that the momentum rate in our simulations
(Figure~\ref{dotlbol}) roughly follows the same trend as the best fit to
the observational data in \citet{2004A&A...426..503W}. However, while our points from the
higher-mass protostellar simulations are within the uncertainty range
from \citet{2004A&A...426..503W}, the points from the lower-mass
simulations are a factor a few above.  
Still, our simulations only follow the evolutionary track of one example massive protostar forming under one clump environmental mass surface density. Most lower luminosity sources in the observational samples are expected to be protostars forming from lower-mass prestellar cores, and may also be in systematically different environments. A proper comparison here will require simulating a broader range of prestellar core masses and environmental conditions and then sampling the core mass function to build a realistic population of protostars at different evolutionary stages.

In addition, we note that the data in
\citet{2004A&A...426..503W} has a large scatter of about two orders of
magnitude on either side of the best fit line, and therefore even our
points from the lower-mass protostellar simulations are in agreement with
some of their data points.  Also, the luminosity of a source depends on the
viewing angle, as much of the luminosity from the accretion and the
protostar will escape out through the outflow cavity, which can lead to large uncertainty in the measurement.  There is also
uncertainty related to the conversion of $^{12}\mathrm{CO}$ to outflow
mass \citep{2016ApJ...832..158Z}, which can lead to the observed
momentum rate being underestimated.  Hence it is possible that even
our low protostellar mass simulations are in better agreement with the
best fit from \citet{2004A&A...426..503W} than what it appears from
Figure~\ref{dotlbol}.  Compared with \citet{2004A&A...426..503W},
\citet{2015MNRAS.453..645M} found a very similar best fit line to
their data, but have somewhat smaller scatter of the data.

Other observational constraints can be made by comparing magnetic-field
strengths in our simulations with observed values. The magnetic field
in massive cores from which massive stars form was measured by
\citet{2018A&A...614A..64B}. They found field strengths of
$\sim0.6-3.7~{\rm mG}$ in the high-mass starless region IRDC 18310-4.
Our initial setup, with a core scale mG magnetic field, is in
reasonable agreement with these findings. Then for
later stages, \citet{2010MNRAS.404..134V} found that in high density
material in Cep A where the masers occur,
the magnetic field strength is $\sim 23~{\rm mG}$. 
Observations have found a field strength of $\sim10~{\rm mG}$ in high 
density regions near the disk in the high-mass protostar IRAS 18089
\citep{2008A&A...484..773V,2010ApJ...724L.113B,2017A&A...607A.111D}.
We find magnetic field strengths at the base of the
outflow approaching 100 mG, in approximate agreement with the findings from 
Cep A,
and the high density and near disk region of IRAS 18089.
\citet{2009A&A...506..757S} and \citet{2017A&A...607A.111D} also found 
that the small
scale field probed by the masers are consistent with large scale
fields traced by dust. 

\subsection{Summary}

In summary, we have performed 3 dimensional magneto-hydrodynamic
simulations of outflows from protostars for a sequence of protostellar
models forming from a $60~{M_\odot}$ prestellar core in a clump environment with mass surface density of 1~g~cm$^{-2}$.  We have found that the
outflow generally becomes stronger and wider as the protostar grows in mass. The evolution of the outflow opening angle (Figure~\ref{openinganglecomp}) agrees well with the semi-analytic model of \citet{2014ApJ...788..166Z}.
The mass
flow rates, momentum flow rates, and outflow masses in our simulations are
in reasonable qualitative agreement with observations \citep{2004A&A...426..503W,2015MNRAS.453..645M}.  With these
simulations, and this particular mass configuration, we find a star
formation efficiency of $\sim50\%$, which is also in good agreement with the
$\sim43\%$ found by the analytic calculations performed by
\citet{2014ApJ...788..166Z}.

\acknowledgements

We thank the referee for helpful comments and suggestions.
Part of this work was performed when J. E. S. was a post doc at the
University of Florida.  This work has been supported in part by NASA
grant NNX15AP95A, in part by NASA through grant HST-AR-15053 from the
Space Telescope Science Institute, which is operated by AURA, Inc.,
under NASA contract NAS 5-26555, and in part by the University of
Florida astro theory post doctoral fellowship program.  This work was
supported in part by NAOJ ALMA Scientific Research Grant number
2017-05A, and JSPS KAKENHI Grant Numbers 19K14760, JP19H05080.  
J.C.T. acknowledges support from NSF grant AST1411527 and ERC project 
788829 —MSTAR.
We would like to thank Y. Zhang for helpful comments.

\bibliography{ms}
\bibliographystyle{aasjournal}
\end{document}